\documentclass{aastex63}

\usepackage{natbib}
\usepackage{rotating}
\usepackage{xcolor}

\accepted{February 2022}

\shorttitle{Nearby M Dwarf Orbit Mapping with SOAR}
\shortauthors{Vrijmoet et al.}

\begin{document}

\title{The Solar Neighborhood XLIX: New Discoveries and Orbits of M
  Dwarf Multiples with Speckle Interferometry at SOAR}

\correspondingauthor{Eliot Halley Vrijmoet}
\email{vrijmoet@astro.gsu.edu}

\author[0000-0002-1864-6120]{Eliot Halley Vrijmoet}
\altaffiliation{Visiting Astronomer, Cerro Tololo Inter-American
  Observatory. CTIO is operated by AURA, Inc., under contract to the
  National Science Foundation.}
\affiliation{Department of Physics and Astronomy,
Georgia State University,
Atlanta, GA 30303, USA}
\affiliation{RECONS Institute, Chambersburg, PA 17201, USA}

\author{Andrei Tokovinin}
\affiliation{Cerro Tololo Inter-American Observatory | NSF's NOIRLab,
  Casilla 603, La Serena, Chile}

\author[0000-0002-9061-2865]{Todd J.\ Henry}
\altaffiliation{Visiting Astronomer, Cerro Tololo Inter-American
  Observatory. CTIO is operated by AURA, Inc., under contract to the
  National Science Foundation.}
\affiliation{RECONS Institute, Chambersburg, PA 17201, USA}

\author[0000-0001-6031-9513]{Jennifer G.\ Winters}
\altaffiliation{Visiting Astronomer, Cerro Tololo Inter-American
  Observatory. CTIO is operated by AURA, Inc., under contract to the
  National Science Foundation.}
\affiliation{Center for Astrophysics | Harvard \& Smithsonian, 60
  Garden Street, Cambridge, MA 02138, USA}

\author[0000-0003-2159-1463]{Elliott Horch}
\affiliation{Department of Physics, Southern Connecticut State
  University, 501 Crescent Street, New Haven, CT 06515, USA}

\author[0000-0003-0193-2187]{Wei-Chun Jao}
\altaffiliation{Visiting Astronomer, Cerro Tololo Inter-American
  Observatory. CTIO is operated by AURA, Inc., under contract to the
  National Science Foundation.}
\affiliation{Department of Physics and Astronomy, Georgia State
  University, Atlanta, GA 30303, USA}

\begin{abstract}

We present the first results of a multi-year program to map the orbits
of M dwarf multiples within 25 parsecs.  The observations were
conducted primarily during 2019--2020 using speckle interferometry at
the Southern Astrophysical Research (SOAR) Telescope in Chile, using
the High-Resolution Camera mounted on the adaptive optics module
(HRCam+SAM).  The sample of nearby M dwarfs is drawn from three
sources: multiples from the RECONS long-term astrometric monitoring
program at the SMARTS 0.9\,m, known multiples for which these new
observations will enable or improve orbit fits, and candidate
multiples flagged by their astrometric fits in \textit{Gaia} Data
Release 2 (DR2).  We surveyed 333 of our 338 M dwarfs via 830 speckle
observations, detecting companions for 63\% of the stars.  Most
notably, this includes new companions for 76\% in the subset selected
from \textit{Gaia} DR2. 
In all, we report the first direct detections of 97 new stellar companions to the observed M dwarfs. 
Here we present the properties of those
detections, the limits of each non-detection, and five orbits with
periods 0.67--29 yr already observed as part of this program.
Companions detected have projected separations of
$0\farcs024$--$2\farcs0$ (0.25--66 AU) from their primaries and have
$\Delta I \lesssim 5.0$ mag.  This multi-year campaign will ultimately
map complete orbits for nearby M dwarfs with periods up to 3 yr, and
provide key epochs to stretch orbital determinations for binaries to
30 yr.
\end{abstract}


\section{Introduction} 
\label{sec:intro}

Stars in binary and multiple star systems have been observed in many
varieties of orbits, each the result of the stellar formation and
dynamical evolution processes that guided them through to the present
day.  Multiples may form from fragmentation at overdensities in the
collapsing molecular cloud \citep{Pri89}, creating gravitationally
bound stars separated by thousands of AU \citep{Off16,Lee19,Kuf19}, or
may form later from the fragmentation of the disk around a (single)
protostar, generating stars separated by 50--200 AU
\citep{Bon94,Kra10}.
Observers, however, have noted a wealth of systems with separations of
$\lesssim$10 AU, indicating that many of these multiples undergo
significant dissipative processes to lose their angular momentum
\citep{Duc13}.  As reviewed in \citet{Bat15} and \citet{Lee20}, such
processes could involve close encounters with nearby stellar neighbors
or interactions with the circumstellar or circumbinary disk(s), such
as accretion, which in turn is affected by magnetic field interactions
and metallicity \citep{Moe19}. 

Clarifying the roles of these
processes requires detailed numerical models and, above all, observed
distributions of the orbital parameters such as orbital period,
semimajor axis, eccentricity, and mass ratio, that are affected by
these dissipative processes. 
For example, a distribution favoring high eccentricities suggests a thermal distribution of orbital velocities produced by dominating dynamical interactions \citep{Kro08}, and has been observed for systems with early-type primary stars \citep{Moe17}. 
Or, as a broader example, if the presence of a disk generally dampens eccentricity, then any trends of eccentricity with semimajor axis could be linked to disk size scales. 
Key information will come especially
from the inclusion of higher-order multiples such as triples and
quadruples, rather than binaries alone, as those systems
carry additional evidence through their ratios of masses and orbital
periods and the mutual inclinations of their orbits.

Previous efforts establishing orbital element distributions for main
sequence multiples have focused on specific spectral type or mass
regimes.  For example, binaries of solar-type stars of types FGK were
the focus of \citet{Duq91} and a succeeding effort by \citet{Rag10}.
Results for early-type binaries with O primary stars were
presented by \citet{Mas98}, with additional analysis that compared the O and B massive stars
to the solar-type stars by \citet{Moe17}.  Each of these efforts has
discussed the observed distribution of eccentricity as a function of
orbital period ($P_\mathrm{orb}$ vs.\ $e$), highlighting that
solar-type and more massive systems show a clear correlation between
period and eccentricity, with the shortest-period systems almost
exclusively circular.  In contrast, the very low-mass systems
($\lesssim$0.1 M$_\odot$) presented by \citet{Dup17} did not show this
correlation.  This result suggests a mass-dependent or age-dependent
difference in dynamical histories or formation pathways of stellar
multiples.

M dwarfs make up $\sim$75\% of all stars \citep{Hen06,Hen18}, and a
detailed study of their orbital architectures would complete the sweep
of stars along the main sequence.  With masses spanning
0.08--0.62~M$_\odot$ \citep{Ben16}, they are the primary product of
the star formation process, so their ubiquity renders their orbital
parameter distributions of particular interest.  In an initial effort,
M dwarf systems showed a solar-type $P_\mathrm{orb}$ vs.\ $e$
distribution in \citet{Udr00}, but their results were limited by their
small sample of 48 systems, and an expanded sample is needed.

To bolster the statistics for M dwarf multiples, we are assembling a
sample of at least 120 M dwarf systems with accurately measured
orbits spanning periods 0--30 yr and semimajor axes up to $\sim$10 AU
(depending on stellar mass). 
This sample size makes this study the largest on M dwarf multiples' orbits to date. 
With a particular focus on orbital
eccentricity, our goals include determining the period at which tidal
circularization occurs and to reveal any structures in the
$P_\mathrm{orb}$ vs.\ $e$ diagram. 
Our specific goal is to determine 120 orbits in an attempt to populate the final $P_\mathrm{orb}$ vs.\ $e$ plot with roughly 20 orbits in each 5-year bin of $P_\mathrm{orb}$, making the eccentricity distributions clear overall as well as within each of those regimes.
The specific goal of 120 orbits has been set to maximize the detail of the final distribution with consideration for availability of resources. 
We are collecting these orbits
from broader sets of multiples observed in the long-term RECONS
(REsearch Consortium On Nearby Stars, \href{http://www.recons.org}{www.recons.org}) astrometry
program \citep[as described in][]{Vri20}, known orbits in the
literature (including the $\sim$30 published from the \citet{Udr00} sample described above, and a new multi-epoch speckle interferometry campaign.

This paper presents the first results of the speckle observations,
which are being carried out at the Southern Astrophysical Research
(SOAR) 4.3\,m telescope in Chile using the High-Resolution Camera
(HRCam) and SOAR Adaptive Optics Module \citep[SAM;][]{Tok18b}.  This
productive telescope-instrument combination has been used to derive
hundreds of high-quality orbits over the past decade
\citep[e.g.,][]{Tok19,Tok20}.  Observations for this M dwarf project
have progressed at a rapid pace since commencing in 2019, with orbital
motion clearly visible already for several targets.  The resulting
characterization of M dwarf multiples, in parallel with our
complementary multiplicity study of K dwarfs \citep{Hen21}, will
provide key comparisons between the lowest mass stars and their
higher-mass cousins, as well as a data set well-suited to constraining
formation and dynamical evolution models of multi-star systems.  In
this paper, we focus on the M dwarfs, describing the sample in
$\S$\ref{sec:sample}, the speckle observations in
$\S$\ref{sec:observations}, and results of the SOAR effort in
$\S$\ref{sec:results}.  Discussion of the results proceeds in
$\S$\ref{sec:discussion}.

%
\section{Sample} 
\label{sec:sample}
%

The targets in this program are 338 known and candidate M dwarf
multiples within 25 pc visible from the Southern Hemisphere.  By the end
of 2020, 333 of these targets have been observed at SOAR.

Distances were determined via parallaxes from the RECONS astrometry
program at the SMARTS 0.9\,m \citep[$\S$\ref{sec:pb_sample} in this paper; also][]{Jao05,Hen18} and \textit{Gaia} DR2
\citep{GAI16,GAI18b}; all systems meet the 25 pc cutoff in one or both
of these catalogs\footnote{A few systems do not meet the 25 pc
  distance cutoff using updated parallaxes from \textit{Gaia} EDR3
  \citep{GAI20b}, which was released after this SOAR program began.}.
The full sample will be volume-limited, but does not need to be
volume-complete.  M dwarfs have been selected as having $V-K_s > 3.70$
using Johnson $V$ and 2MASS $K_s$ (hereafter $K$) filters, as well as
absolute magnitude $M_V > 9.02$. 
These limits were established as the $M_V$ and $M_K$ values corresponding to 0.6 M$_\odot$ using the \citet{Ben16} mass-luminosity relation for M dwarfs. 
This sample thus spans spectral types M0 through M9. 
 For 11 systems that had no $V$
measurements available, we converted the \textit{Gaia} DR2 $B_G$ and
$R_G$ magnitudes to $V$ using the relations for M dwarfs in
\citet{Jao18}.  Finally, the specifications of HRCam+SAM on SOAR limit
the sample to systems brighter than $I = 14$ mag and south of
$+$25$^\circ$ in declination.


The primary goal of the project is to map the distribution of orbital
eccentricity with respect to orbital period, with the sample of 338
systems intended to support an even representation of periods 0--30
yr.  Although determining 120 accurate orbits is the primary goal, the
speckle sample includes several times that many systems; this reflects
our expectation that only a subset will have well-defined orbits with
$P_\mathrm{orb} < 30$ yr by the end of the 3-year observing campaign.
To reach 120 orbits, the full project sample will include orbits
observed using additional methods from other programs with a variety
of time baselines and strengths, e.g., long-term astrometry and
systems with spectroscopic orbits.  Because this paper presents
results of the speckle subset only, hereafter the ``sample'' and
similar terms will refer to the speckle subsample rather than the
ultimate full project sample that will include all observing methods.

Table~\ref{tab:table_targets_lookup} lists the entire speckle sample
of 338 M dwarfs targeted at SOAR, including the five stars not yet
observed by the end of 2020.  For each target are listed Right
Ascension and Declination 2000.0 positions (columns 1--2), the
WDS-style coordinate name (column 3), the WDS discoverer code if the
pair has been previously resolved (column 4), and the target name used
in other RECONS work (column 5).  These identifying parameters are
followed by each system's parallax in milliarcseconds (mas; column 6)
and the reference for that value (column 7), the $V$ magnitude and
reference (columns 8 and 9), and the $V-K$ color (column 10), where
$K$ is from 2MASS \citep{2MASS}.  Given next are the subsets to which each target
belongs (columns 11--13, described in detail below) and flags (column
14) for whether the system has been resolved (Y) or not resolved (N)
thus far at SOAR (N/A indicates not yet observed), with the flag
``T2'' marking systems with results presented in
Table~\ref{tab:table_targets_results}.  Finally, a reference for the
orbit of a system is given (column 15), if it exists, with flag ``T4''
in this column marking systems with orbits presented in this work
($\S$\ref{sec:orbits}).

\begin{figure}[h!]
\epsscale{0.48}
\plotone{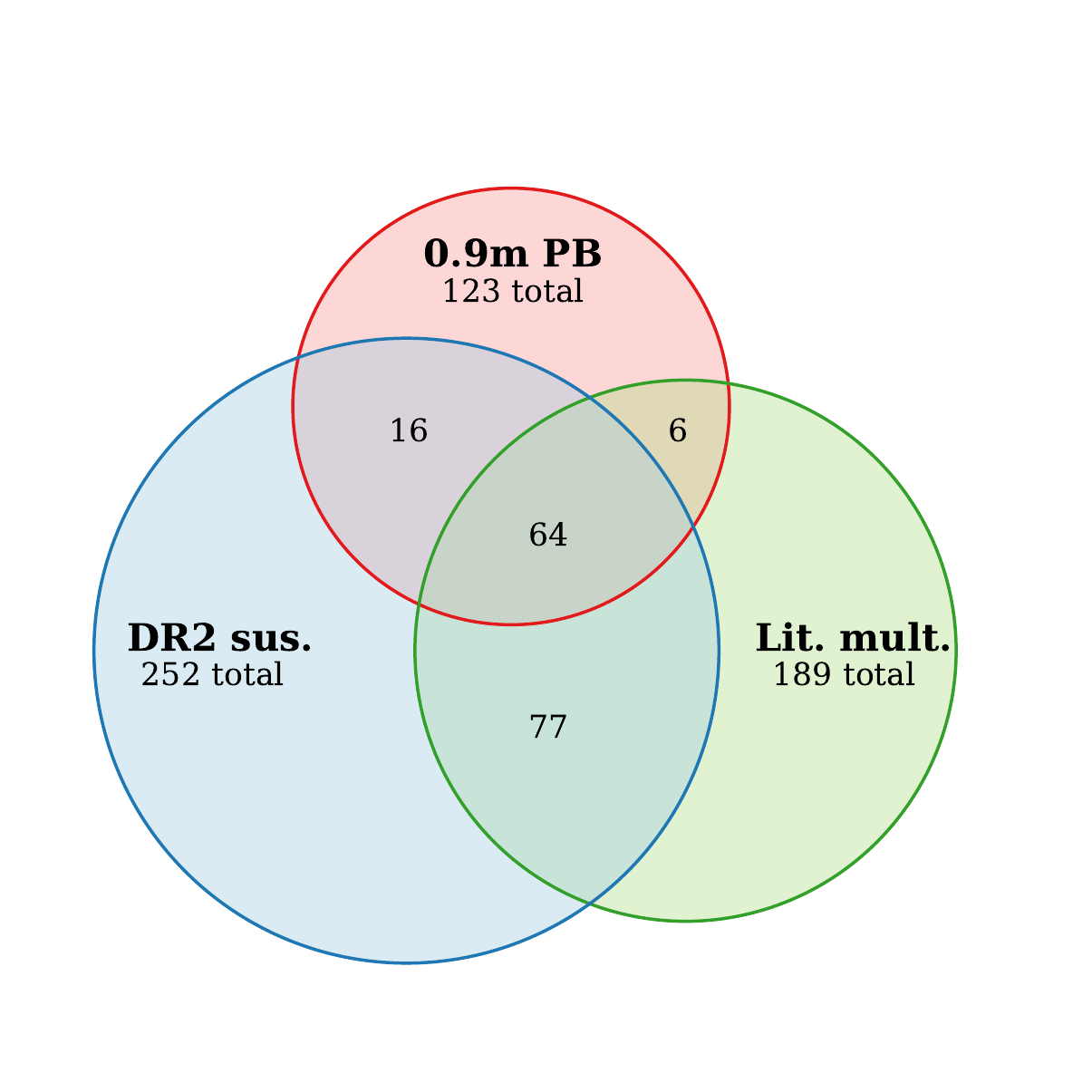}
\vspace{-2em}
\caption{Venn diagram illustrating the three subsets of the SOAR
  sample of nearby M dwarfs.  The area of each circle is proportional
  to the number of targets in that subset, but the overlapping regions
  are not to scale.  Each circle is labeled with a reference to the
  subset's source: ``0.9\,m PB'' for targets showing perturbations
  (PBs) in the RECONS astrometry program at the CTIO 0.9\,m,
  ``Lit.\ mult.'' for known literature multiples, and ``DR2 sus.'' for
  systems suspected to be multiples based on their \textit{Gaia} DR2
  results.  The number of targets is given under each subset name, and
  the numbers in the overlapping sections indicate the number of
  targets common to multiple subsets.
\label{fig:samplevenn}}
\end{figure}

The target list of 338 systems is drawn from three sources:
astrometric multiples identified through long-term RECONS data
\citep{Jao05,Hen18}, known multiples from the literature with
potential $P_\mathrm{orb} < 30$ yr, and suspected multiples chosen
based on their \textit{Gaia} DR2 results \citep[criteria described
  in][]{Vri20}.  As illustrated in Figure~\ref{fig:samplevenn}, these
subsets overlap each other --- for example, some systems from RECONS
astrometry are already known multiples in the literature --- and in
the target list in Table~\ref{tab:table_targets_lookup} we have
indicated each target's subset membership using columns 11--13.  The
selection and goals for each of these groups is described next.

\subsection{123 Targets from the RECONS Astrometry Program}
\label{sec:pb_sample}

The RECONS program \citep{Jao05,Hen18} began taking astrometry data in
1999, targeting red, brown, and white dwarfs within 25 pc.  Through
4--6 observing runs per year at the SMARTS 0.9\,m at CTIO, this
program has been mapping the motions of several hundred nearby stars
for a median duration of 10 yr.  This enables the detection of
binaries with orbital periods many decades in length, with orbital
characterization possible for $P_\mathrm{orb}$ up to $\sim$30 yr in
the longest-observed cases.  Fully observed orbits are fit using the
Markov Chain Monte Carlo method introduced in \citet{Die18}, which
simultaneously fits the proper motion, parallax, and orbital motion of
the system's photocenter; nine examples with $P_\mathrm{orb}$ from
2--17 yr using RECONS data from the 0.9\,m are given in \citet{Vri20}.


RECONS astrometry qualified the selection of 123 targets for the SOAR
observing list, with 37 of these targets not qualifying for either of
the other two subsets.  Systems were considered high priority if their
residuals to the parallax and proper motion fit exhibited
perturbations (PBs) that were characteristic of orbital motion due to
bound companions.  
These residuals are considered significant perturbations if their maximum 
amplitude is at least three times the size of the 
average error per epoch for that system (with these errors typically 
3--5 mas). 
In many cases these residuals clearly traced out orbital motion by the 
system's photocenter, with a smooth rise and fall in R.A.\ 
and/or Decl.\ axes, depending on orbital coverage, observing cadence, 
and the particular orbit shape. 
Orbital period can be estimated by eye in these cases, or constrained by 
a preliminary fit to an astrometric orbit model. 
Targets were selected for our SOAR speckle campaign
if these residuals thus indicated an orbit with likely $P_\mathrm{orb}
\lesssim$ 30 yr. In other select cases the residuals were clearly perturbed
but the motion was more difficult to interpret, which may occur when
an orbit shorter than $\sim$3 yr is observed with the relatively
sparse cadence of the RECONS observations or the PB is weak because
two components have similar fluxes and the photocenter consequently
moves very little.

The goals for the ``0.9\,m PB'' subset (column 11 in
Table~\ref{tab:table_targets_lookup}) are thus twofold:

\begin{itemize}

\item For systems with orbits that can be fully characterized in the
  RECONS astrometry, resolving the components will allow us to
  determine their individual dynamical masses \citep[following the
    methods outlined in][]{vdK67}. \\[-0.6cm]
\item For targets with PBs that are ambiguous rather than clearly due
  to orbital motion, resolving a second star will confirm that
  companion and constrain its orbit, aiding interpretation of the
  RECONS astrometric residuals and ongoing observing priorities for
  the 0.9\,m program. 

\end{itemize}

In both cases, non-detections will place constraints on the natures of
the potential companions and their orbits, and in some cases (notably,
in the unclear ones) non-detections will allow us to rule out a
companion as the source of the astrometric residuals.

\subsection{189 Targets from Known Multiples in the Literature}
\label{sec:kno_sample}

To enrich the sample, and because astrometry is less sensitive to some
types of binaries (e.g., equal luminosity components), the SOAR target
list was augmented with known M dwarf multiples from the literature.
These known multiples constitute 189 targets, with 42 not belonging to
either of the other subsets.  Our observations are intended to capture
orbital motion, so these targets were limited to pairs that had
previously been resolved at separations $\lesssim 2\farcs0$ or likely
orbital periods less than 30 yr.  Not all of these pairs have been
resolved in the literature; about a third are known multiples based on
only spectroscopic or astrometric results.  These systems were
primarily selected by cross-matching the Sixth Catalog of Orbits of
Visual Binary Stars \citep{ORB6} against coordinates of M dwarfs from
\textit{Gaia} DR2 and the RECONS astrometry target list.  These were
augmented by some M dwarf multiples from the Washington Double Star
Catalog \citep[WDS;][]{WDS} and private communications from
collaborators.

The intention of the observations for this ``Literature multiples''
subset (column 12 of Table~\ref{tab:table_targets_lookup}) is to add
new measurements to the existing data sets for each system, with the
following goals:

\begin{itemize}

\item Enable fitting of each system's relative orbit by extending the
  time baseline of observations. \\[-0.6cm]

\item Improve upon any existing orbit fits, in particular by refining
  the precision of the orbital elements. \\[-0.6cm]

\end{itemize}



\startlongtable
\begin{longrotatetable}
\begin{deluxetable}{cccllrcccccccll}
\tabletypesize{\scriptsize}
\tablecaption{Target list for the SOAR speckle program for 25 pc M dwarfs.
\label{tab:table_targets_lookup}}
\tablehead{
\colhead{R.A.   } & \colhead{Decl.  } & \colhead{WDS} &  \colhead{Discov.} &\colhead{Name} & \colhead{$\pi$} & \colhead{$\pi$} & \colhead{$V$  } & \colhead{$V$ } & \colhead{$V - K$} &  \colhead{0.9m} & \colhead{Lit. } & \colhead{DR2   } & \colhead{SOAR} & \colhead{Orbit} \\[-1em] 
\colhead{J2000.0} & \colhead{J2000.0} & \colhead{   } &  \colhead{code   } &\colhead{    } & \colhead{(mas)} & \colhead{ref.}  & \colhead{(mag)} & \colhead{ref.} & \colhead{(mag)  } &  \colhead{PB  } & \colhead{mult.} & \colhead{sus.} & \colhead{res.} & \colhead{ref. } \\[-1em] 
\colhead{(1)    } & \colhead{(2)    } & \colhead{(3)$^a$} &  \colhead{(4)    } &\colhead{(5) } & \colhead{(6)  } & \colhead{(7)$^b$}   & \colhead{(8)  } & \colhead{(9)} & \colhead{(10)$^c$} &  \colhead{(11)$^d$} & \colhead{(12)$^d$} & \colhead{(13)$^d$} & \colhead{(14)$^e$} & \colhead{(15)$^f$} \\[-2em] 
}
\startdata
00 06 39.24 & $-$07 05 35.9 & 00067$-$0706 & JNN 11           & 2MA 0006-0705 AB       & $46.960  \pm 0.403 $ & EDR3  & 14.72  & APdr9  &  5.76  &                & $\checkmark$   & $\checkmark$   & N, T2 &        \\ 
00 08 53.92 & $+$20 50 25.6 & 00089$+$2050 & BEU 1            & G 131-026 AB           & $55.256  \pm 0.761 $ & DR2   & 13.52  & Rie14  &  5.51  & $\checkmark$   & $\checkmark$   & $\checkmark$   & Y   & T3     \\ 
00 09 45.04 & $-$42 01 39.3 & 00098$-$4202 &                  & LEHPM 1-0255 AB        & $60.889  \pm 0.350 $ & EDR3  & 13.62  & Win15  &  5.40  &                &                & $\checkmark$   & Y, T2 &        \\ 
00 13 46.60 & $-$04 57 37.2 & 00138$-$0458 &                  & LHS 1042              & $42.627  \pm 0.219 $ & EDR3  & 17.98  & estim  &  7.50  &                &                & $\checkmark$   & N, T2 &        \\ 
00 15 27.99 & $-$16 08 01.8 & 00155$-$1608 & HEI 299          & GJ 1005 AB             & $169.522 \pm 0.969 $ & Vri20 & 11.48  & Win15  &  5.09  & $\checkmark$   & $\checkmark$   & $\checkmark$   & Y   & Ben16  \\ 
00 15 58.07 & $-$16 36 57.6 & 00160$-$1637 & BWL 2            & 2MA 0015-1636 AB       & $56.096  \pm 0.093 $ & EDR3  & 13.20  & Win19  &  5.29  &                & $\checkmark$   & $\checkmark$   & Y   & T3     \\ 
00 16 01.97 & $-$48 15 39.1 & 00160$-$4816 & TOK 808          & L 290-072 AB           & $40.672  \pm 0.525 $ & EDR3  & 11.55  & Koe10  &  4.44  &                & $\checkmark$   & $\checkmark$   & Y   &        \\ 
00 16 14.63 & $+$19 51 37.5 & 00162$+$1952 &                  & GJ 1006 AC             & $65.108  \pm 0.041 $ & EDR3  & 12.26  & Wei96  &  5.17  &                & $\checkmark$   & $\checkmark$   & Y, T2 &        \\ 
00 21 37.26 & $-$46 05 33.4 & 00216$-$4606 &                  & L290-028             & $51.569  \pm 0.045 $ & EDR3  & 12.24  & Koe10  &  4.79  &                &                & $\checkmark$   & N, T2 &        \\ 
00 24 44.19 & $-$27 08 24.2 & 00247$-$2653 & LEI 1AB          & GJ2005 AB             & $129.317 \pm 0.126 $ & EDR3  & 15.28  & Win15  &  7.04  & $\checkmark$   & $\checkmark$   & $\checkmark$   & Y   & Koe12  \\ 
00 24 44.10 & $-$27 08 24.0 & 00247$-$2653 & LEI 1BC          & GJ2005 BC             & $129.317 \pm 0.126 $ & EDR3  & 15.28  & Win15  &  7.04  & $\checkmark$   & $\checkmark$   & $\checkmark$   & Y   & Man19  \\ 
00 25 04.31 & $-$36 46 17.9 & 00251$-$3646 & BRG 2            & LTT00220 AB           & $49.871  \pm 0.110 $ & EDR3  & 12.48  & Win15  &  4.65  &                & $\checkmark$   & $\checkmark$   & Y   &        \\ 
\enddata

\tablerefs{
* = This work, 
APdr9 = \citet{APdr9}, 
And07 = \citet{And07}, 
Ben16 = \citet{Ben16}, Bur15b = \citet{Bur15}, 
Cal17 = \citet{Cal17}, 
Dah88 = \citet{Dah88}, Dit14 = \citet{Dit14}, Doc19 = \citet{Doc19}, Dup10b = \citet{Dup10}, Dup16 = \citet{Dup16}, 
For99 = \citet{For99}, 
EDR3 = \textit{Gaia} EDR3 \citep{GAI20b}, DR2 = \textit{Gaia} DR2 \citep{GAI18b}, 
Hei94 = \citet{Hei94}, 
Hen18 = \citet{Hen18}, HIP07 = \citet{HIP07}, 
Izm19 = \citet{Izm19}, 
Jao14 = \citet{Jao14}, 
Ker16 = \citet{Ker16}, Koe10 = \citet{Koe10}, Koe12 = \citet{Koe12}, Kon10 = \citet{Kon10}, 
Lur14 = \citet{Lur14}, 
Man19 = \citet{Man19}, Mas18 = \citet{Mas18}, 
Rie10 = \citet{Rie10}, Rie14 = \citet{Rie14}, Rie18 = \citet{Rie18}, 
Seg00 = \citet{Seg00}, Sca19 = \citet{Sca19}, Sod99 = \citet{Sod99}, 
Tok15c = \citet{Tok15}, Tok17b = \citet{Tok17}, Tok18a = \citet{Tok18a}, Tok18c = \citet{Tok18c}, Tok19c = \citet{Tok19}, Tok19b = \citet{Tok19b}, Tok20a = \citet{Tok20a}, Tok20b = \citet{Tok20}, 
Vri20 = \citet{Vri20}, 
Wei96 = \citet{Wei96}, 
Win15 = \citet{Win15}, Win17 = \citet{Win17}, Win19 = \citet{Win19}, 
Zir03 = \citet{Zir03}
}
\tablenotetext{a}{Column 3 --- For all systems not already noted in
  the WDS catalog \citep{WDS}, the WDS code given is the anticipated
  code for the future entry should these systems be resolved.}
\tablenotetext{b}{Column 7 --- The parallax reference is ``EDR3'' for
  \textit{Gaia} EDR3 \citep{GAI20b}, ``DR2'' for DR2 \citep{GAI18b},
  or other references as listed in the Table notes.}
\tablenotetext{c}{Column 10 --- Reference for all $K$ magnitudes in
  $V-K$ color is \citep{2MASS}.}
\tablenotetext{d}{Columns 11--13 --- These are classification flags
  indicating the subsets to which each system belongs: 0.9\,m PB =
  system with perturbation in RECONS astrometry residuals
  ($\S$\ref{sec:pb_sample}), Lit.\ mult.\ = known binary from the
  literature ($\S$\ref{sec:kno_sample}), DR2 sus.\ = system with
  evidence of multiplicity in \textit{Gaia} DR2 results
  ($\S$\ref{sec:gai_sample}).  Check marks in parentheses
  ($\checkmark$) in column 12 indicate unpublished results from
  coauthor Winters (to be published speckle survey results).}
\tablenotetext{e}{Column~14 --- This column indicates SOAR resolutions
  and non-resolutions as presented in previous papers in the yearly
  SOAR series \citep[e.g.,][]{Tok19,Tok20,Tok21}, except those noted
  with the ``T2'' flag, which are given in
  Table~\ref{tab:table_targets_results}.}
\tablenotetext{f}{Column~15 --- This column gives the reference for
  the existing orbit in the literature from the Sixth Catalog of
  Orbits of Visual Binary Stars \citep{ORB6}, with flag ``T4'' noting
  orbits newly presented in this work in Table~4 and Figure~2.}
\tablecomments{This table is available in its entirety in machine-readable form.}
\end{deluxetable}
\end{longrotatetable}

\subsection{252 Targets Selected from Gaia DR2}
\label{sec:gai_sample}

\textit{Gaia} DR2 \citep{GAI16,GAI18b} released proper motions and
parallaxes for $\sim$1.7 billion sources based on an astrometric model
that includes only those two sources of motion, with orbital motion
fits not planned until future data releases.  Systems exhibiting
orbital motion from a bound companion should thus exhibit evidence of
poor astrometric fits.  \cite{Vri20} showed that nearby M dwarfs with
unresolved companions can be selected based on several DR2 fit
parameters, akin to the astrometric residuals in RECONS data
($\S$\ref{sec:pb_sample}).

\textit{Gaia} DR2 results were used to identify 252 total M dwarfs for
the SOAR observing list, with DR2 being the only source of potential
multiplicity for 95 targets.  This evidence is based on the analysis
of \citet{Vri20}, and most of these ``DR2 suspects'' met at least some
of the final criteria presented there.  
Those specific DR2 criteria identified in \citet{Vri20} were:
\begin{enumerate}
\item missing parallax or missing catalog entry, 
\item \texttt{parallax\_err} $\geq 0.32$ mas for $G \lesssim 18$ ($\geq 0.40$ mas otherwise), 
\item \texttt{astrometric\_gof\_al} $\geq 56.0$, 
\item \texttt{astrometric\_excess\_noise} $\geq 108.0$, and
\item \texttt{ruwe} $\geq 2.0$.
\end{enumerate}
That work found that at least three out of four systems meeting at least one of these thresholds were multiples unresolved in DR2. 
While selecting targets for
this subset of SOAR observations, we anticipated that the values of these criteria may
eventually be lowered if many stars that were presumed single are
later revealed to be binary. 

The goals for this group of ``DR2 suspects''(column 13 of
Table~\ref{tab:table_targets_lookup}) are:

\begin{itemize}

\item Map orbits of new multiples with periods that will be at least
  50\% complete by the end of this 3-year observing campaign (i.e.,
  with $P_\mathrm{orb} \lesssim$ 6 yr). The DR2 selection criteria
  should be more sensitive to these particular systems because of its
  relatively short observing baseline of 22 months.

\item Confirm the validity of the \citet{Vri20} criteria for selecting
  binaries from \textit{Gaia} DR2 via the resolution of companions,
  and revise the criteria if necessary.

\end{itemize}

%
\section{Observations and Data Reduction with HRCam+SAM} 
\label{sec:observations}
%

The observations presented here were made over 2018--2020, with most
completed between July 2019 and December 2020, representing the first
half of our planned 3-year program.  Many systems in our sample were
already observed at SOAR prior to this project as part of earlier
initiatives to investigate M dwarfs in the Southern Hemisphere.  Their
results do not appear in Table~\ref{tab:table_targets_results} because
those results were presented in previous SOAR papers \citep{Tok21,Tok20}; instead, they have
a ``Y'' or ``N'' in column~13 of Table~\ref{tab:table_targets_lookup}
with no additional flags.



Time awarded for the speckle observing programs of coauthors Tokovinin
and Vrijmoet was combined in order to increase the opportunities for timely
observations of fast-orbiting systems.  In preparation for each
observing run, previous SOAR observations and RECONS astrometry were considered, and systems that
had exhibited rapid orbital motion were prioritized for the upcoming
run.  This procedure improved the likelihood that defining features of
the orbit shapes would not be missed.

All of the observations used HRCam, the high-resolution camera
mounted on the SOAR Adaptive Module \citep[SAM,][]{Tok16}, in the
seeing-limited mode (no laser guide star was used).  Frames were taken
almost exclusively in the Kron-Cousins $I$ filter, usually in 2--3
sets (data cubes) of 400 frames per target, with integrations
typically 24 ms per frame.  These sets were each later processed
independently to verify results.  Most observations use the HRCam
narrow $3^{\prime\prime}$ field of 200$\times$200 pixels, whereas
pairs known to have separations of $1\farcs 4$ or more were observed
with the 400$\times$400 field.  The resolution limit in $I$ is usually
40--45 mas depending on target brightness and sky conditions, but can
be as close as 35 mas in some cases \citep[see Figure 1 of][]{Tok20}.
Targets that are unresolved in the first two attempts are usually
observed a third time, then retired from the program if still
unresolved.


\startlongtable
\begin{deluxetable}{ccchcccccccc}
\tabletypesize{\scriptsize}
\tablecaption{Results of observations through 2020 in the SOAR speckle program for 25 pc M dwarfs. 
All magnitude differences are in the $I$ band, except where the $y$ band is noted in column 11. 
\label{tab:table_targets_results}}
\tablehead{
\colhead{WDS} & \colhead{First} & \colhead{Date obs.} & \nocolhead{Filt.} & \colhead{Resol.} & \colhead{$\rho$             } & \colhead{$\theta$} & \colhead{$\Delta m$} & \colhead{$\rho_\mathrm{min}$} & \colhead{$\Delta m$ ($0\farcs15$)} & \colhead{$\Delta m$ ($1\farcs0$)} & \colhead{Obs. }  \\[-1em] 
\colhead{   } & \colhead{res. } & \colhead{(year)   } & \nocolhead{     } & \colhead{(Y/N) } & \colhead{($^{\prime\prime}$)} & \colhead{(deg)   } & \colhead{(mag)     } & \colhead{($^{\prime\prime}$) } & \colhead{(mag)                    } & \colhead{(mag)                   } & \colhead{flags}  \\[-1em] 
\colhead{(1)$^a$} & \colhead{(2)$^b$} & \colhead{(3)      } & \nocolhead{(4)  } & \colhead{(4)   } & \colhead{(5)$^c$            } & \colhead{(6)$^c$ } & \colhead{(7)$^c$  } & \colhead{(8)$^d$            } & \colhead{(9)$^d$                  } & \colhead{(10)$^d$                } & \colhead{(11)$^e$}  \\[-2em] 
}
\startdata
00067$-$0706    & Jan14 & 2019.8568 & $I  $ & N &        &       &     & 0.0768 & 2.3 & 2.9 &       \\ 
                &       & 2020.8342 & $I  $ & N &        &       &     & 0.0594 & 2.3 & 2.8 &       \\ 
00098$-$4202    &   *   & 2019.6133 & $I  $ & Y & 0.0522 & 159.0 & 0.8 &        &     &     &       \\ 
                &       & 2019.8567 & $I  $ & N &        &       &     & 0.0525 & 2.5 & 3.9 &       \\ 
                &       & 2020.8341 & $I  $ & Y & 0.0959 & 115.7 & 1.0 &        &     &     &       \\ 
                &       & 2020.9270 & $I  $ & Y & 0.1089 & 117.4 & 0.9 &        &     &     & q     \\ 
00138$-$0458    & none  & 2019.8568 & $I  $ & N &        &       &     & 0.1145 & 1.6 & 2.5 & :     \\ 
                &       & 2020.8342 & $I  $ & N &        &       &     & 0.1260 & 1.6 & 1.6 & :     \\ 
00162$+$1952    &   *   & 2019.5397 & $I  $ & N &        &       &     & 0.0636 & 2.3 & 3.9 &       \\ 
                &       & 2019.8564 & $I  $ & N &        &       &     & 0.0543 & 2.7 & 4.1 &       \\ 
                &       & 2020.9241 & $I  $ & Y & 0.0312 & 44.8  & 1.0 &        &     &     &       \\ 
\enddata

\tablerefs{
Jan14 = \citet{Jan14}, 
Jao14 = \citet{Jao14}, 
Jod13 = \citet{Jod13}, 
Kar20 = \citet{Kar20}, 
Mar00 = \citet{Mar00}, 
War15 = \citet{War15}
}
\tablenotetext{a}{Column 1 --- For resolved systems not already noted
  in the WDS catalog \citep{WDS}, the WDS code given is the
  anticipated code for the future entry.}
\tablenotetext{b}{Column 2 --- This column gives a single or double
  asterisk (* or **) for each new resolution, depending on previous
  status of the target's multiplicity.  A single asterisk (*)
  indicates a new resolution of a system already known in the
  literature to be a multiple, but which has never previously been
  resolved.  A double asterisk (**) marks a new resolution of a system
  that was previously a multiple candidate at best, with its
  multiplicity not established in the literature; these are new
  multiples.  Systems previously resolved by others have their first
  resolution reference listed.  Systems not resolved here and never
  resolved previously are noted with ``none'' in this column.}
\tablenotetext{c}{Columns 5--7 --- For observations that resolved a
  companion, these columns give the separation ($\rho$), position
  angle ($\theta$), and magnitude difference ($\Delta m$) between
  components.}
\tablenotetext{d}{Columns 8--10 --- For observations with no detected
  companion, these columns provide limits: the minimum separation
  distinguishable ($\rho_\mathrm{min}$) for pairs with $\Delta m < 1$
  mag, the magnitude difference limit at $0\farcs15$ from the primary
  source, and the magnitude difference limit at $1\farcs0$ from the
  source.}
\tablenotetext{e}{Column 11 --- This column contains flags related to
  each observation: q = quadrant has been determined, p = $\Delta m$
  determined photometrically from average image, : = noisy data, $y$ =
  magnitude difference in $y$ band (all others in $I$ band).}
\tablecomments{This table is available in its entirety in machine-readable form.}
\end{deluxetable}

The data are processed and reduced for this program using the standard
procedures described in \citet{Tok10} and \citet{Tok18b}, and representative images of the reduced data products are shown in \citet{Tok18b}.  In brief,
for each target the power spectrum and autocorrelation function are
calculated, and companions are noted via power spectrum fringes or
secondary peaks in the autocorrelation function.  Fitting an empirical
model to the power spectrum yields the parameters of each detected
pair: the separation between components ($\rho$), the position angle
of the secondary with respect to primary star ($\theta$) (north =
0$^\circ$ through east = 90$^\circ$), and the difference in magnitude
between components ($\Delta m$).  Important details about these
results are:

\begin{itemize}

\item The position angle determined through this procedure is only
  ascertainable modulo 180$^\circ$, leaving some ambiguity in the
  secondary's true position on the sky. 
  This ambiguity has been eliminated whenever possible by applying a shift-and-add procedure to each target's data \citep{Tok18b}; this process reveals the true quadrant for companions that are not too faint but still have some magnitude difference with their primary star ($\Delta m \gtrsim 0$ mag). 
  These results are noted with the ``q''
  flag in Table~\ref{tab:table_targets_results}, indicating that the
  quadrant has been determined.

\item For some observations of wider pairs a separate procedure is used to determine
  the magnitude difference using the average image for a target
  \citep[described in detail in][]{Tok10}. 
  This method produces more reliable photometry for these cases where the stars' separations are greater than image resolution, reducing bias from speckle anisoplanatism. 
  Observations with $\Delta m$
  determined with this method are marked by a ``p'' in
  Table~\ref{tab:table_targets_results}, indicating that this
  photometric method has been used.

\item For observations in which no companion was detected, a contrast
  curve is computed to report the detection (magnitude) limits as a
  function of the distance from primary star on the sky \citep[for
    example, see Figure~5 of][]{Tok18b}.  The parameters of this curve
  are reported in the results in Table~\ref{tab:table_targets_results}
  as the minimum separation resolvable for pairs with $\Delta m < 1$
  mag, as determined from the maximum spatial frequency of the power
  spectrum, and the maximum detectable magnitude difference at
  separations of $0\farcs15$ and $1\farcs0$ (the dynamic range).

\end{itemize}

%
\section{Results} 
\label{sec:results}
%

Through the end of 2020 and including previously published results,
333 targets on this program have been observed at least once at SOAR
via 830 total observations.  Of these targets, 211 (63\% of the total
sample) had a companion detected at least once, representing 204 total
systems\footnote{In seven cases, a higher-order multiple with two
  components separated by a few arcseconds represents two targets for
  this speckle survey, and as such is represented by two lines in
  Table~\ref{tab:table_targets_lookup} (and counts as two targets
  throughout this paper).}.  
  In this first half of our 3-year program,
most companions were observed numerous times to confirm that the
detected object was a true companion and not a background source; the
remainder have follow-up observations planned.  For each true
multiple, these initial observations will then contribute to that
system's orbit mapping.

The results are detailed in Table~\ref{tab:table_targets_results} for
both newly resolved and unresolved systems.  Targets with previous
resolutions appear instead in the yearly SOAR publication series
\citep[e.g.,][]{Tok20,Tok21}.  Table~\ref{tab:table_targets_results}
gives the WDS coordinate name or anticipated WDS name in column 1.  In
column 2 is either the reference for the first resolution of that
system, a single asterisk (*) for the first resolution of a known
multiple, two asterisks (**) for the first resolution of a system that
was previously, at best, only a candidate multiple (see
$\S$\ref{sec:detections} for details), or ``none'' if the system was
not resolved.  Each observation of a target is then distinguished by
its date (column 3) and Y/N flag for whether or not the companion was
detected at that epoch (column 4).  Observations in which the
companion was resolved include the separation, position angle, and
magnitude difference between components (columns 5--7).  Observations
with no detected companion list the minimum resolution detectable and
$\Delta m$ limits at $0 \farcs 15$ and $1 \farcs 0$ from the primary
star, respectively (columns 8--10).  Finally, observation flags
(column 11) note several of the cases described in
$\S$\ref{sec:observations}, such as when the quadrant of the position
angle is unambiguously determined (q), when the magnitude difference
was determined photometrically from the average image (p), when the
observations resulted in noisy data (:), generally leading to less
robust limits, or $y$ for the one observation done through a $y$
filter rather than the $I$ filter.

Uncertainties on the individual measurements are not listed here, as these would require a more detailed analysis than feasible for this paper. 
The full measurement errors consist of internal errors, which could be determined by comparing each observation's data cubes, and external errors, which can be estimated from HRCam measurements of well-characterized binaries (``calibrators''), \textit{Gaia} resolved sources \citep{Tok19}, and residuals of each system's orbital fit. 
The typical deviation of the calibrators from their orbit models is 1--3 mas in separation and 0.2$^\circ$ in position angle, and in a similar procedure with SOAR speckle data, \citet{Man19} found errors of 3.8 mas and 0.94$^\circ$ are appropriate additions to the internal errors (typically $\leq$2 mas). 
For this reason we have assigned errors of 5 mas to all SOAR HRCam measurements when fitting orbits, and postponed the full derivation of external errors until this 3-year observing program is complete. 
See $\S$\ref{sec:orbits} for additional details of the orbit fitting routine.


\subsection{Detections}
\label{sec:detections}


Table~\ref{tab:summary} summarizes the detection rates for each group
within the full sample ($\S$\ref{sec:sample}).  For each named subset
(column 1), it provides the number of targets observed (column 2), the
number resolved (column 3), the percentage of observed targets that
were resolved (column 4), and the number of targets not yet observed
by the end of 2020 (column 5).

\begin{deluxetable}{lcccc}[h!]
\tablecaption{Summary of SOAR speckle results for each of the three sample subsets, as well as the targets meeting the formal multiplicity criteria in DR2 \citep{Vri20} and the full sample. 
\label{tab:summary}}
\tablehead{
\colhead{Subset} & \colhead{Targets } & \colhead{Pairs   } & \colhead{Percent } & \colhead{Targets not} \\[-1em]
\colhead{name  } & \colhead{observed} & \colhead{resolved} & \colhead{resolved} & \colhead{observed   } \\[-1em]
\colhead{(1)   } & \colhead{(2)     } & \colhead{(3)     } & \colhead{(4)     } & \colhead{(5)        }  \\[-2em]
}
\startdata
0.9\,m PB                   & 120 &  59 & 49\% & 3 \\
Literature multiples        & 188 & 140 & 74\% & 1 \\
DR2 suspects                & 249 & 188 & 76\% & 3 \\
\quad 2$+$ DR2 criteria     & 217 & 176 & 81\% & 2 \\
\tableline
Full sample$^a$             & 333 & 211 & 63\% & 5
\enddata
%
\tablenotetext{a}{Numbers are not the sums of the four categories
  above because of overlaps in samples, as shown in the Venn diagram
  of Figure~\ref{fig:samplevenn}.}
\end{deluxetable}

Of the 211 companions resolved in our sample, 97 had no previously
published resolutions, making these results their first published
positional measurements.  These newly resolved systems are marked with
asterisks in column 2 of Table~\ref{tab:table_targets_results}, broken
into two categories.  A single asterisk (*) denotes the 34 systems
that were already reported to be multiples based on other published
data, e.g., astrometry or spectroscopy.  A double asterisk (**)
denotes new resolutions for 63 systems with no previously reported
multiplicity in the literature --- these were included in the target
list due to anomalies in their RECONS or \textit{Gaia} DR2 astrometry.
These are newly discovered multiples in addition to being new
resolutions.
    
Additionally, 114 companions noted here as resolved at SOAR already
had resolutions in the literature; nearly all of these systems are
listed as ``Y'' in Table~\ref{tab:table_targets_lookup} but without
the ``T2'' flag, as they are presented in \citet{Tok20},
\citet{Tok21}, and previous publications in that yearly series.
Column~2 of Table~\ref{tab:table_targets_results} gives the reference
for the first resolution of that system.  For all systems with data
already in the literature, the new observations presented here and in
the other SOAR results papers will ultimately be combined with
previous results to improve orbital coverage.  We have already
employed this strategy for the orbits we are presenting here
($\S$\ref{sec:orbits}).

Data from the RECONS astrometry program at the SMARTS 0.9\,m already
reveals perturbations in 59 of the 211
resolved pairs.  That astrometry provides maps of the photocentric
orbits, hence the resolutions of companions in these cases will enable
us to solve for the individual masses within each pair as in, e.g.,
\cite{Die18}.  Each of these new masses will contribute to the
currently modest number of dynamically determined individual M dwarf
masses known to date \citep{Ben16}.


Finally, there are 249 targets observed (and 3 targets not observed)
that showed some evidence of poor astrometric fits in \textit{Gaia}
DR2 and were included based on preliminary results of the
\citet{Vri20} analysis.  Our SOAR observations reveal that 188 (76\%)
of these M dwarfs host a companion.  This result highlights the
utility of that method of selecting likely multiples using
\textit{Gaia}'s astrometric fit parameters, especially for these
nearby, low mass systems.  See $\S$\ref{sec:gaiaanalysis} for further
discussion of this result, details about the DR2 criteria outlined in
Table~\ref{tab:table_targets_results}, and the implications.

\begin{figure}[h!]
\epsscale{0.8}
\plotone{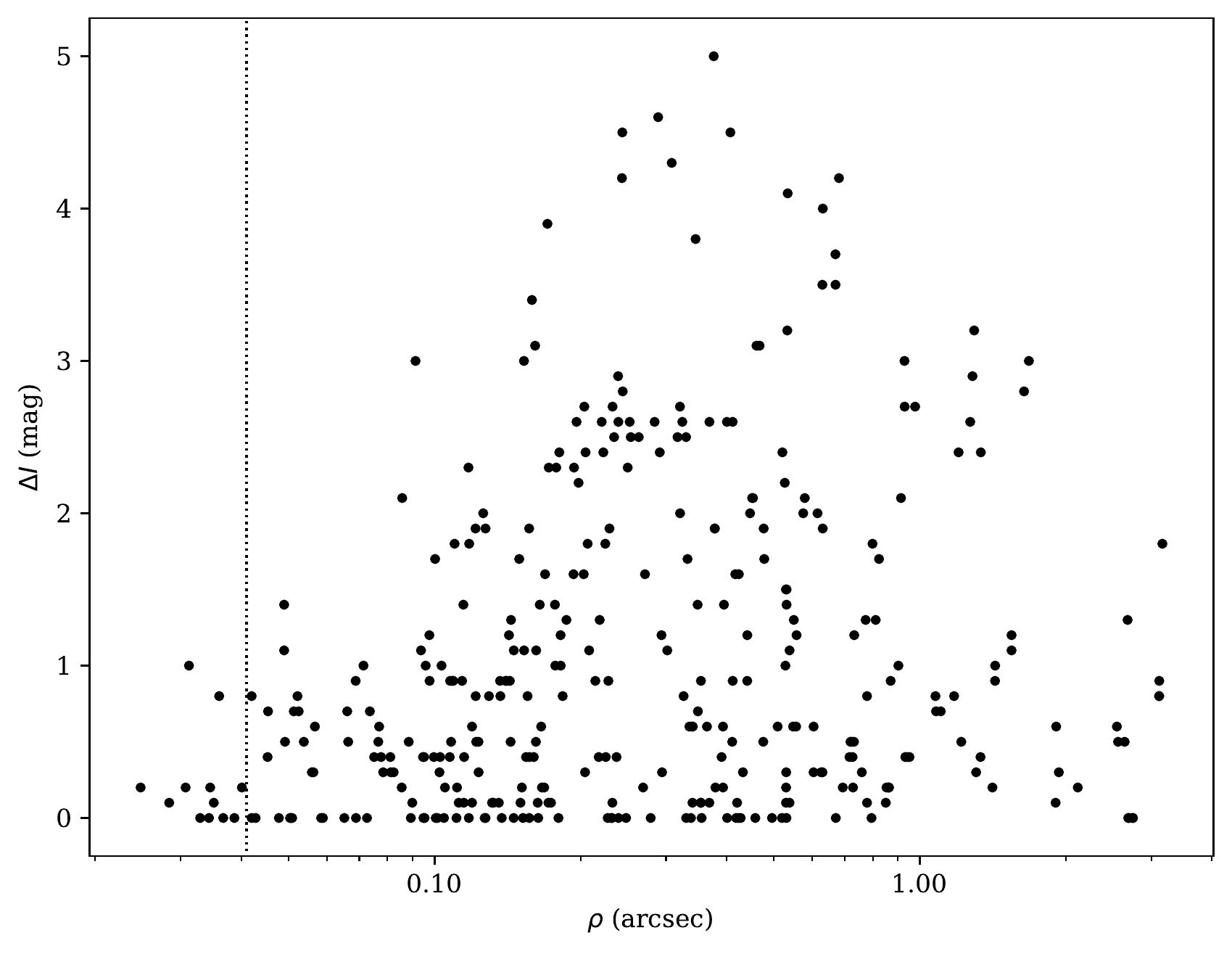}
\caption{Separation $\rho$ in arcsec and magnitude difference in $I$ band for each observation that resolved a companion, excluding those for which the data were exceptionally noisy (``:'' flag in Table~\ref{tab:table_targets_results}). The 41~mas formal diffraction limit of SOAR is indicated with the vertical dotted line. This sample is intentionally focused on the closer pairs ($\lesssim 1\farcs0$) that are more likely to show orbital motion over our 3-year campaign. 
\label{fig:seps-dmags}}
\end{figure}

Figure~\ref{fig:seps-dmags} shows the separations ($\rho$) and magnitude differences in $I$ band ($\Delta I$) for each observation that detected a companion. 
This distribution of exclusively M dwarfs is similar to that of the wider sample observed yearly by SOAR \citep[shown in Figure~1 of][]{Tok20}. 
The most notable difference is our distribution shows a paucity of systems with $\Delta I > 1.5$~mag and $\rho < 0\farcs1$. 
This discrepancy could reflect the higher fraction of very faint companions in our sample compared to the other samples observed yearly at SOAR.
The mass-luminosity relation is known to experience a severe drop at optical wavelengths at low M dwarf masses \citep{Ben16}. Therefore, it is not surprising that companions only slightly less massive than their primaries may have large $\Delta I$ values compared to their primaries and remain undetected at SOAR.

\subsection{Non-detections}
\label{sec:non-detections}

The 122 targets observed with no companions detected at SOAR still
impart important information via the detection limits given in columns
8--10 of Table~\ref{tab:table_targets_results}.  Because these
observations were conducted in the $I$ filter, in many cases these
non-detections restrict potential companions to the regimes of cool
white dwarfs, very low-mass stars, or brown dwarfs.  Examples include
LHS 1582 AB (03434$-$0934), SCR 0723-8015 AB (07240$-$8015), and LP
848-50 AB (10427$-$2416), all of which exhibit clear orbital motion in
their long-term astrometry \citep[][]{Win17,Vri20}.  Other true
multiples unresolved here may have orbits too tight to resolve, or
have components positioned unluckily too close to each other on the
sky at the epoch of observation.  In each of these cases, the
non-detection information given here provides constraints on orbits
and companion masses that can be used in concert with other efforts to
reveal information about any unseen and/or undetected companions.

\vspace{1cm}
\subsection{Orbits}
\label{sec:orbits}

\begin{figure}[h!]
\epsscale{0.95}
 \plottwo{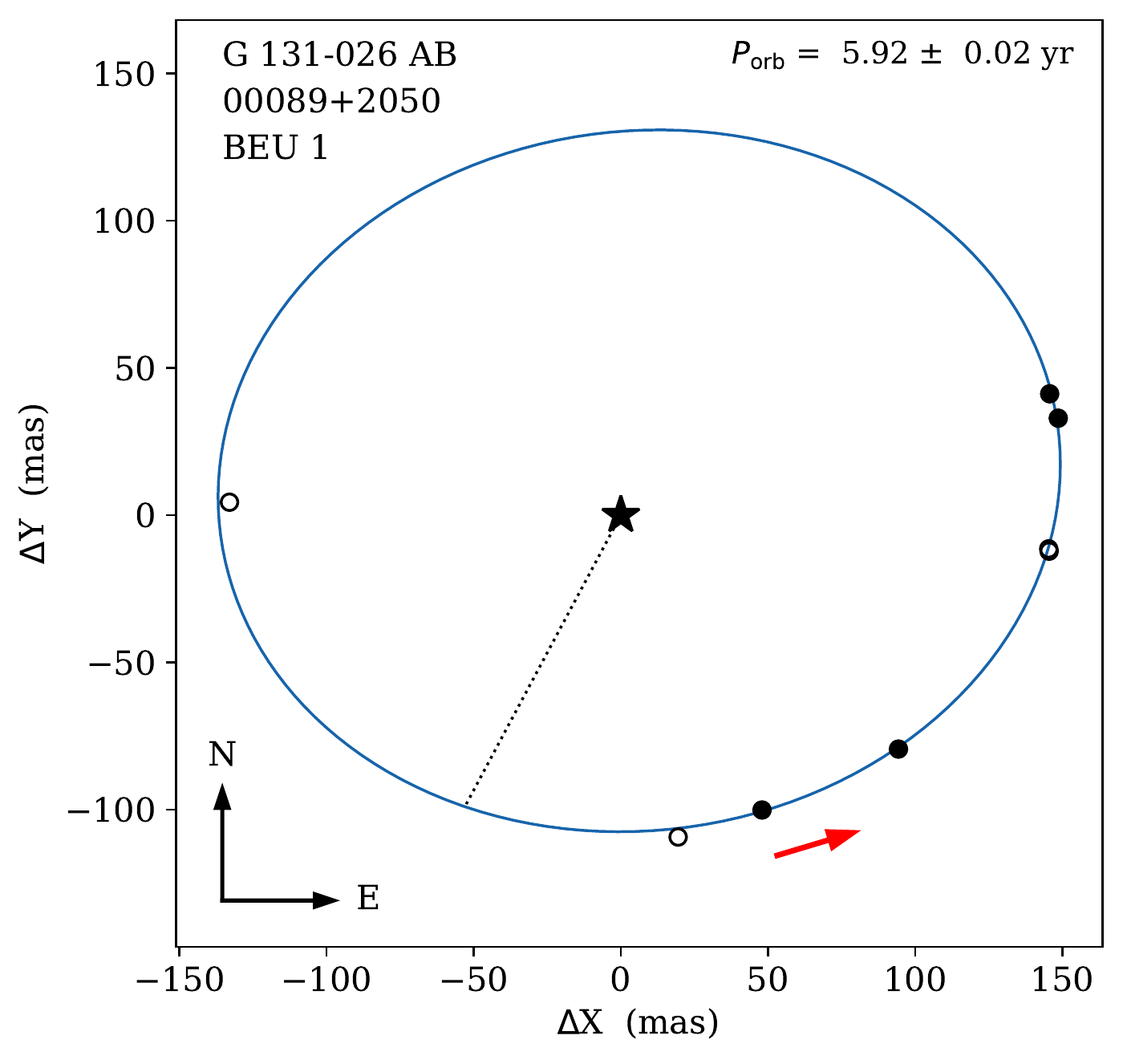}{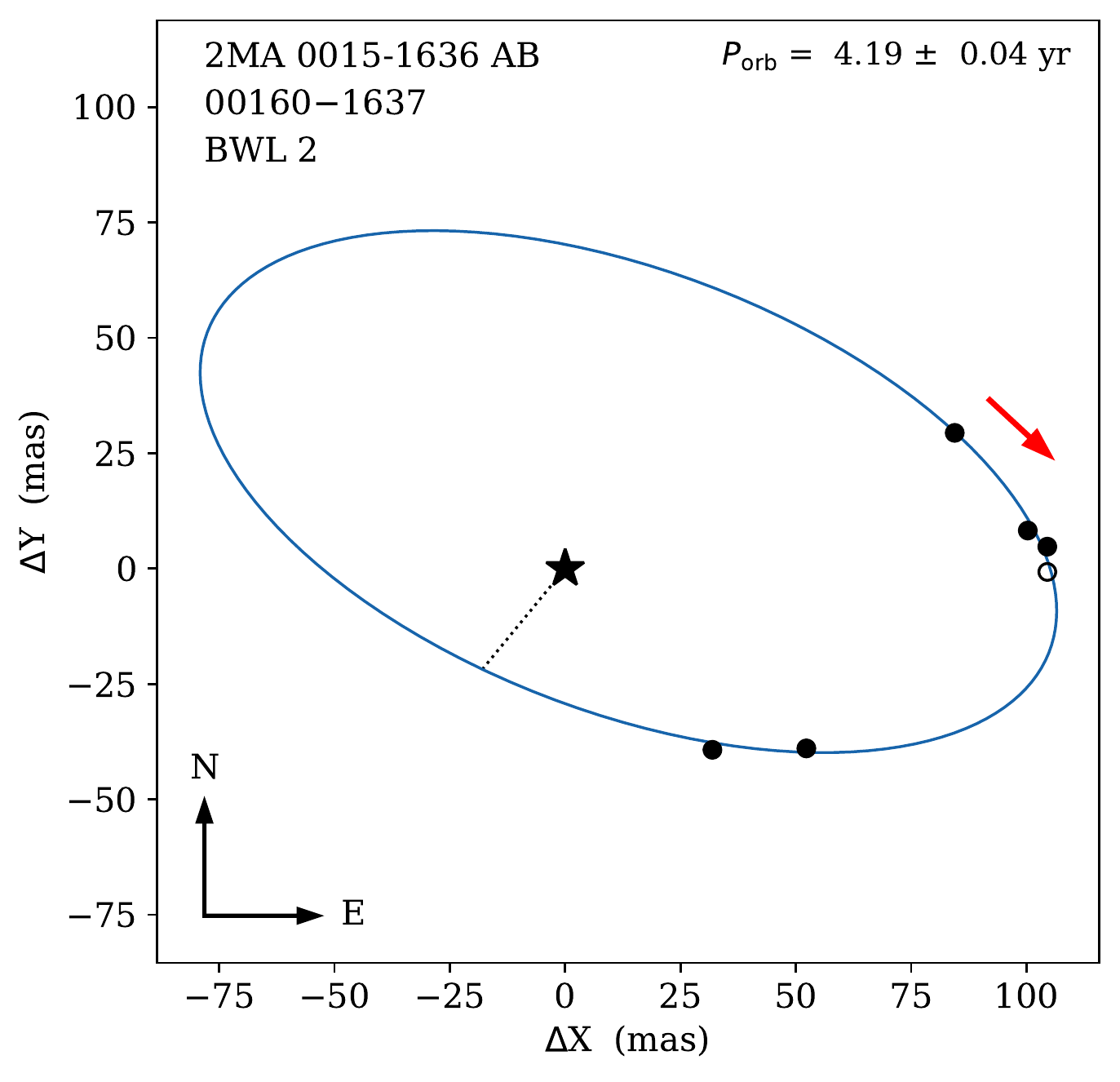}
 \plottwo{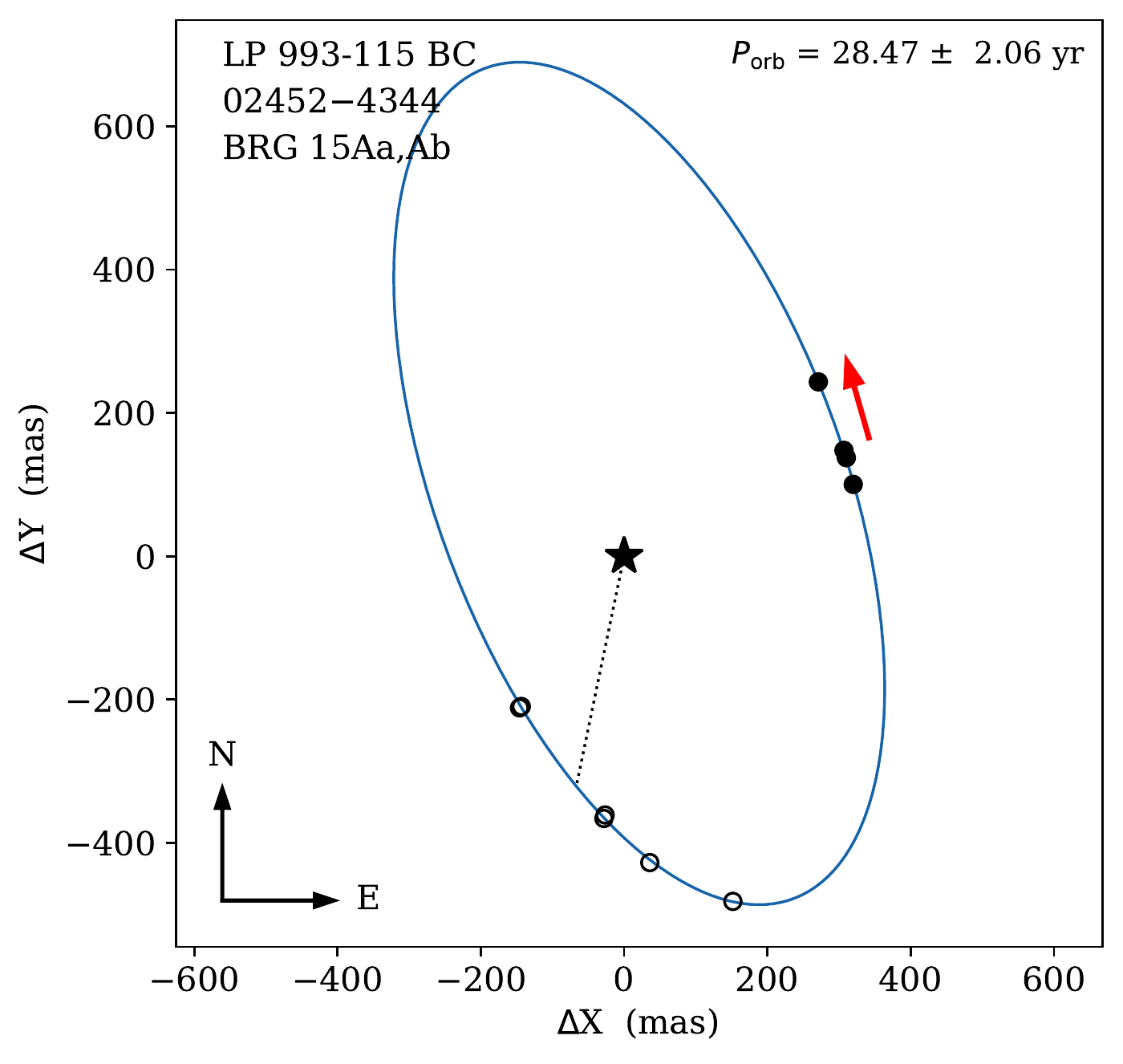}{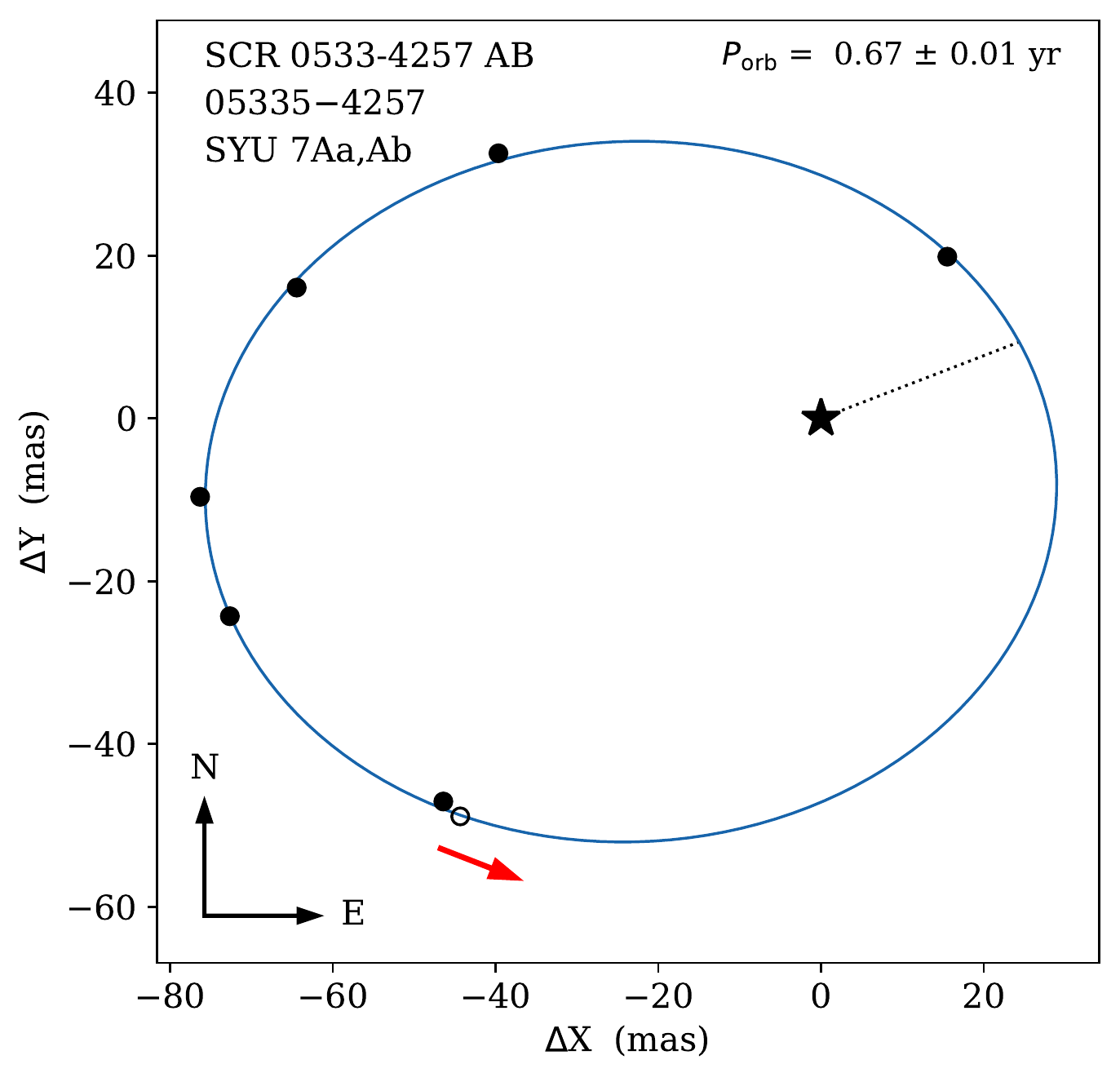}
 \plottwo{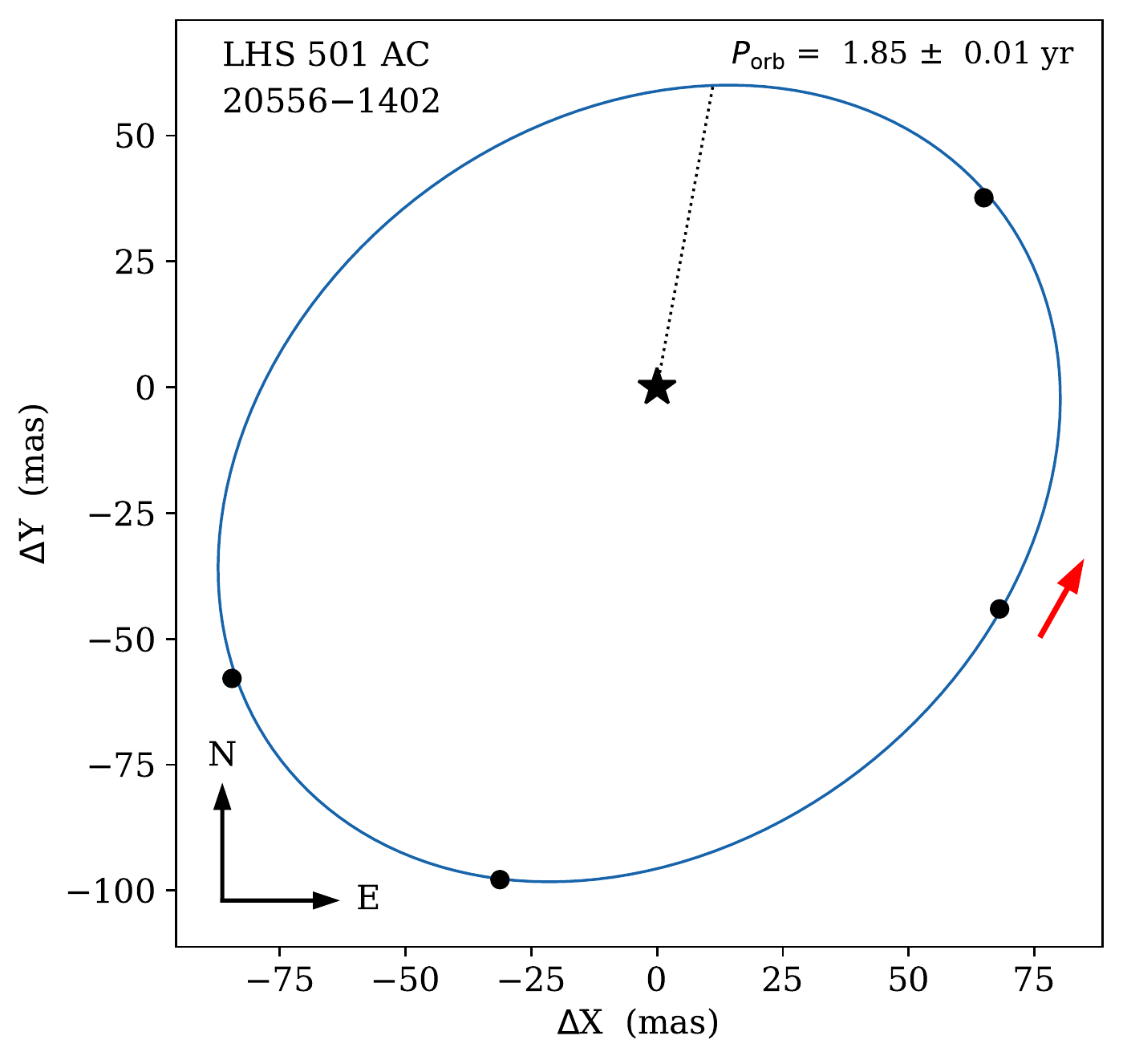}{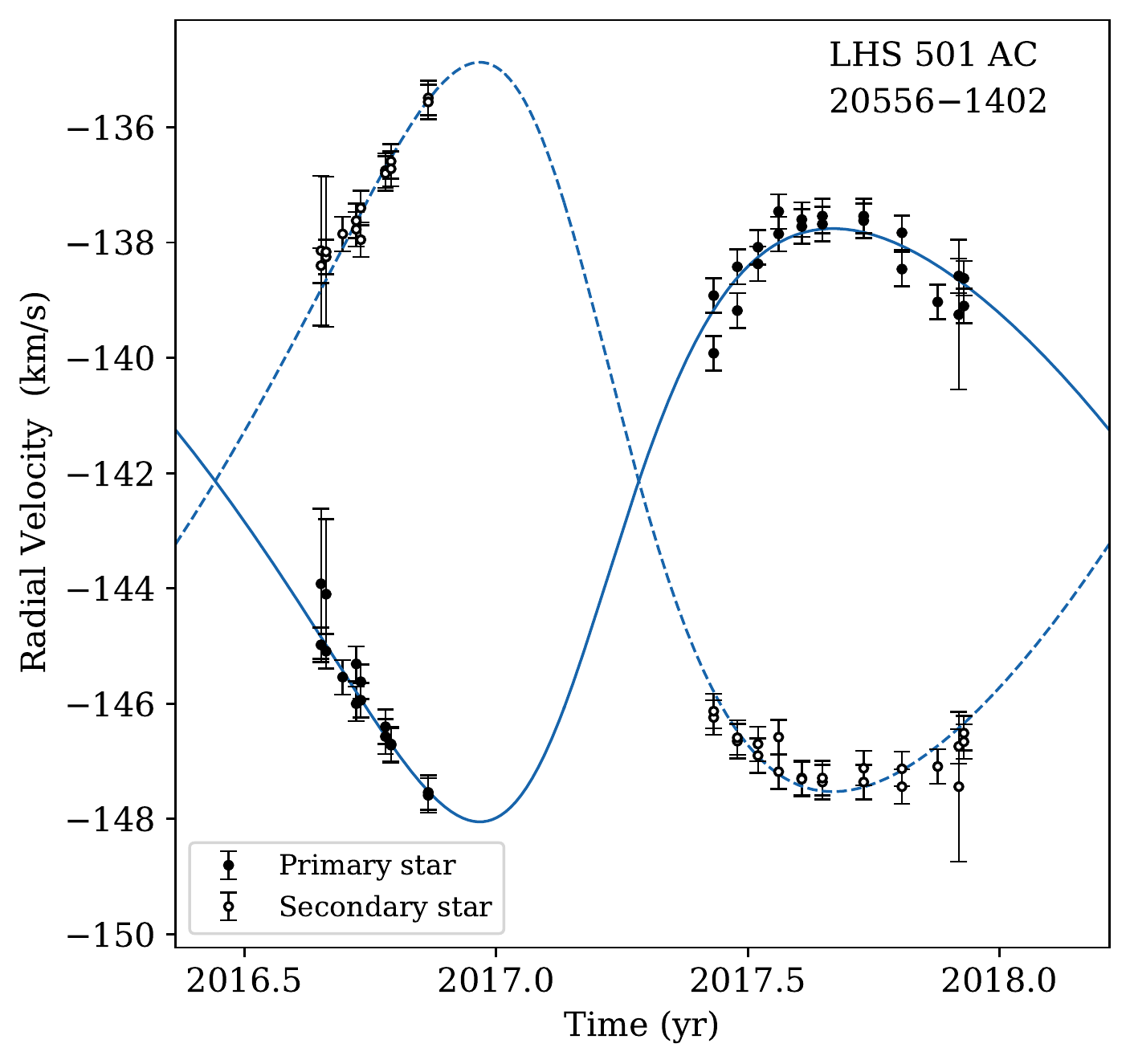}
\caption{Five visual orbits for binaries resolved at SOAR, plus one
  spectroscopic orbit that was fit simultaneously with the
  corresponding visual orbit for LHS~501~AC.  For the visual orbits,
  blue lines denote the fit, filled circles are SOAR observations, and
  open circles are observations added from the literature.  Red arrows
  indicate the direction of motion of the secondary star around the
  primary, and the black star and dotted line denote the primary star
  and line of periastron.  In the spectroscopic orbit (bottom right
  panel), the points and solid line are the observations and fit,
  respectively, for the primary component, and the open points and
  dashed line are the observations and fit for the secondary.
  \textit{Left to right, top to bottom:} G~131-26~AB,
  2MA~0015-1636~AB, LP~993-115~BC, SCR~0533-4257~AB, LHS~501~AC visual
  orbit, and LHS~501~AC radial velocity orbit \citep[observations
    from][]{Bar18}. Sources for additional visual observations are
  specified in Table~\ref{tab:orbitsSOAR}.
\label{fig:prelimorbits}}
\end{figure}

Here we present five orbits fit using the SOAR observations, often
combined with additional data available in the literature; all but LHS
501 AC are the first orbits for the systems.  
These five orbits represent the highest-quality fits possible with the data from this program thus far, and fortuitously are also representative of the range of size and time scales accessible to this program. 
The orbital periods
range from 0.67--29 yr, and each has at least four observations taken
during the first 1.5 years of this observing program.  The full
orbital parameters are given in Table~\ref{tab:orbitsSOAR} and
illustrations of the fits are shown in Figure~\ref{fig:prelimorbits}.
Each dataset was fit with the \texttt{ORBIT} code \citep{Tok16orbit}, which
uses 
the Leavenberg-Marquardt least-squares method to identify the model orbit that best fits the weighted observations. 
The weights are inversely proportional to the errors on each point, which for these observations have been set to the
typical external HRCam errors of 5 mas, and for literature observations are set
to the published errors.  The resulting
fits have errors ranging from 0.3\%--7.2\% in orbital period and
1.3\%--6.7\% in semimajor axis.  
These errors on the orbital parameters are determined by the fitting algorithm. 

Each system with an orbit fit is
discussed briefly below.  In each case we also provide estimates of
the component masses using our work toward a mass-luminosity relation
in $I$ band \citep{Vri21}.  These estimates should be considered
preliminary and are only intended as general guides of the mass
regimes for these M dwarfs.

\begin{itemize}
\item G 131-26 AB (00089$+$2050, BEU 1) is a known flare star with a
  stellar companion first detected by \citet{Beu04} in 2001, then
  resolved again in 2012 by \citet{Jan14} and in 2014 by
  \citet{Hor15}.  We have resolved it four additional times in
  2019--2020 and fit all of these data together to determine an
  orbital period of $5.918 \pm 0.017$ yr.  Combining this orbit with
  the \textit{Gaia} EDR3 parallax indicates a total system mass of
  $0.51 \pm 0.05$ M$_\odot$.  The individual components' $M_I$ values
  are consistent with 0.3 M$_\odot$ and 0.2 M$_\odot$, a good match to
  the system total mass.





\item 2MA 0015-1636 AB (00160$-$1637, BWL 2) was resolved by
  \citet{Bow15} in 2011, who suggested an orbital period of 4.5~yr
  based on their observed separation.  With our additional five points
  we find an orbital period of $4.187 \pm 0.039$ yr, yielding a total
  mass of $0.41 \pm 0.08$ M$_\odot$ using the EDR3 parallax.  The
  individual components' absolute magnitudes imply masses of 0.25--0.3
  M$_\odot$ for each component.  These values are somewhat higher than
  indicated by the total dynamical mass, pointing to some inaccuracy
  in the orbit or the parallax. 
  This is validated by the \textit{Gaia} reduced unit weight error (RUWE) value of 4.1, indicating the parallax is not well fit.  
  The mass discrepancy would be eliminated by a 7\% smaller parallax, or by increasing the orbit's semimajor axis by 7\% or decreasing its period by 10\%. 
  Continued observations at SOAR will allow us to refine the orbit,
  and future \textit{Gaia} data releases will likewise improve the
  system's parallax.



\item LP 993-115 BC (02452$-$4344, BRG 15Aa,Ab), also known as LP
  993-116 AB, is a common proper motion companion to LP 993-115 A
  \citep[44$^{\prime\prime}$;][]{Bid85}.  The C component was first
  identified by \citet{Ber10} with lucky imaging and resolutions also
  reported in \citet{Jan12} and \citet{Jan14}.  We add four new
  resolutions to map the other side of the orbit, and derive a period
  of $28.466 \pm 2.056$ yr.  Using the EDR3 parallax, this suggests a
  total mass of $0.42 \pm 0.23$ M$_\odot$ for BC, although this value
  is poorly constrained.  Individual absolute magnitudes for the B and
  C components are consistent with component masses of 0.2--0.25~M$_\odot$ 
  each.  This is the first orbit published for this
  subsystem of this higher-order multiple.





\item SCR 0533-4257 AB (05335$-$4257, SYU 7Aa,Ab) was first resolved
  by \citet{Sha17} in 2014, and to this we have added six points in
  2019--2020.  With an orbital period $0.672 \pm 0.003$ yr, the orbit
  and EDR3 parallax indicate a total mass of $0.40 \pm 0.07$
  M$_\odot$.  This is consistent with the possible period of 9 months
  noted in the unresolved RECONS astrometry by \citet{Rie18}.  The
  individual absolute magnitudes of each component are consistent with
  0.25 M$_\odot$ and 0.15 M$_\odot$, together an excellent match to
  the total dynamical mass.





\item LHS 501 AC (20556$-$1402) is a now-resolved primary with a wide
  companion known as LHS 500, separated by 107$^{\prime\prime}$
  \citep{Jao03}.  The AC pair had not been resolved prior to this
  work, but was noted to be an astrometric multiple by \citet{Jao11}
  based on the RECONS astrometry data.  \citet{Bar18} noted it to be
  SB2 and presented a spectroscopic orbit fit.  Our new orbit was fit
  to their spectroscopic data simultaneously with our four new visual
  resolutions using the same \texttt{ORBIT} code as for the other 
  four orbits in this work. The resulting orbital period of $1.855 \pm 0.014$
  yr is shorter in length but ten times more precise than the
  \citet{Bar18} period ($2.22 \pm 0.16$ yr).  Our eccentricity is also
  significantly different, at $0.242 \pm 0.008$ vs.\ their $0.402 \pm
  0.059$.  Additional observations underway will significantly improve
  future orbit fits for this system, as the radial velocity model
  still shows some minor discrepancies with the data (lower rightmost
  panel of Figure~\ref{fig:prelimorbits}).  With our result and the
  EDR3 parallax we derive a total mass of $0.37 \pm 0.02$~M$_\odot$.
  The individual absolute magnitudes correspond to stars with masses
  of 0.25 M$_\odot$ and 0.2 M$_\odot$, which is roughly consistent to
  the total dynamical mass, although future refinement will be
  necessary for this orbit.




\end{itemize}

%
\section{Discussion} 
\label{sec:discussion}
%


As outlined in $\S$\ref{sec:intro}, our goal is to catalog at least
120 orbits with $P_\mathrm{orb} \leq 30$ yr with reliably determined
orbital periods and eccentricities through the combination of this
3-year speckle campaign, the long-term RECONS astrometry program at
the SMARTS 0.9\,m, and orbits in the literature.  A set representing
orbits out to (at least) 30-year periods will be necessary to draw
significant conclusions about the formation and evolution of these
systems. 
Selecting 120 orbits evenly distributed in orbital period will ensure that there are $\sim$20 orbits in each 5-year bin of $P_\mathrm{orb}$ in the final $P_\mathrm{orb}$ vs.\ $e$ plot. 
More fundamentally, the goal of 120 orbits is a compromise between the need to characterize the $P_\mathrm{orb}$ vs.\ $e$ relation with maximum detail and a realistic expectation based on our prior experience and availability of resources. 

The abundance of detected companions and promising initial orbit fits
resulting from this first phase of our SOAR effort provide several
advances toward the overall project regarding orbit distributions of M
dwarf multiples.  For all systems in
Table~\ref{tab:table_targets_results}, both the resolutions and
non-resolutions reported here provide valuable constraints on the
orbits of their companions or the likelihood of each star's
multiplicity.  Notably, we have increased the total number of resolved
M dwarf multiples within 25 pc by 97, representing 194 targets for
further study of M dwarf multiples' properties.  We have also secured
observations for 140 systems that had been previously resolved,
providing not only new points for orbit determinations, but relative
fluxes in the $I$ band that can be used for a robust mass-luminosity
relation at $I$.  Finally, the five new orbits presented here can be
added to the key $P_\mathrm{orb}$ vs.\ $e$ plot, and each new fit
helps to identify reliable orbits as well as those for which more data
are required to reach the orbital element precision needed to reveal
clues about the star formation process.

\begin{longrotatetable}
\begin{deluxetable}{lclhhcccccccc}
\tablecaption{\parbox{9in}{Elements of the best-fit relative orbits shown in Figure~\ref{fig:prelimorbits}. These orbits have been fit to the relative positions of the systems' components using all resolutions (including literature and this SOAR program), as well as radial velocities if those data are available.}
\label{tab:orbitsSOAR}}
\tablehead{
\colhead{Name} &  \colhead{WDS} & \colhead{Discov} & \nocolhead{R.A.   } & \nocolhead{Decl.  } &  \colhead{$P$ } & \colhead{$a$  } & \colhead{$e$} & \colhead{$i$  } & \colhead{$\Omega$} & \colhead{$\omega$} & \colhead{$T_0$} & \colhead{Additional}  \\[-1em]
\colhead{    } &  \colhead{   } & \colhead{code  } & \nocolhead{J2000.0} & \nocolhead{J2000.0} &  \colhead{(yr)} & \colhead{(mas)} & \colhead{   } & \colhead{(deg)} & \colhead{(deg)   } & \colhead{(deg)   } & \colhead{(yr) } & \colhead{data used}  \\[-1em]
\colhead{(1) } &  \colhead{(2)} & \colhead{(3)   } & \nocolhead{(4)    } & \nocolhead{(5)    } &  \colhead{(6) } & \colhead{(7)  } & \colhead{(8)} & \colhead{( 9) } & \colhead{(10)    } & \colhead{(11)    } & \colhead{(12) } & \colhead{(13)      }  \\[-2em]
}
\startdata
 G 131-026     AB & 00089$+$2050 & BEU 1       & 00 08 53.92  & $+$20 50 25.6  & $5.918 \pm 0.017$  &  $144.0 \pm 4.6 $ & $0.106 \pm 0.023$ & $145.67 \pm 3.29$ & $ 83.61 \pm 6.59$  & $240.38 \pm 17.17$   & $2018.921 \pm 0.173$ & Beu04, Jan14, Hor15 \\
 2MA 0015-1636 AB & 00160$-$1637 & BWL 2       & 00 15 58.07  & $-$16 36 57.6  & $4.187 \pm 0.039$  &  $108.0 \pm 7.2 $ & $0.433 \pm 0.090$ & $ 63.63 \pm 2.11$ & $111.81 \pm 5.15$  & $ 98.03 \pm  4.82$   & $2021.145 \pm 0.057$ & Bow15 \\
 LP 993-115    BC & 02452$-$4344 & BRG 15Aa,Ab & 02 45 14.32  & $-$43 44 10.6  & $28.466 \pm 2.056$ &  $630.3 \pm 37.5$ & $0.240 \pm 0.029$ & $117.13 \pm 1.98$ & $158.70 \pm 1.01$  & $305.03 \pm  8.74$   & $2009.714 \pm 0.434$ & Ber10, Jan12, Jan14 \\
 SCR 0533-4257 AB & 05335$-$4257 & SYU 7       & 05 33 28.04  & $-$42 57 20.6  & $0.672 \pm 0.003$  &  $ 54.4 \pm 3.3 $ & $0.490 \pm 0.066$ & $150.73 \pm 9.59$ & $109.41 \pm 29.17$ & $ 44.38 \pm  9.59$   & $2017.155 \pm 0.013$ & Sha17 \\
 LHS 501       AC & 20556$-$1402 & ... ... & 20 55 37.74  & $-$14 02 08.1  & $1.855 \pm 0.014$  &  $ 91.6 \pm 1.2 $ & $0.242 \pm 0.008$ & $142.67 \pm 2.11$ & $236.06 \pm 2.46$  & $232.05 \pm  1.88$    & $2017.135 \pm 0.009$ & Bar18\tablenotemark{\footnotesize a}  \\
\enddata

\tablerefs{
Bar18 = \citet{Bar18}, Ber10 = \citet{Ber10}, Beu04 = \citet{Beu04}, Bow15 = \citet{Bow15}, 
Hor15 = \citet{Hor15}, 
Jan12 = \citet{Jan12}, Jan14 = \citet{Jan14}, 
Sha17 = \citet{Sha17}
}
\tablenotetext{a}{RV data used in orbit fit.}
\end{deluxetable}
\end{longrotatetable}

\subsection{Contributions to Nearby M Dwarf Orbits}


As of July 2021, $\sim$200 orbits in the Sixth Catalog of Orbits of
Visual Binary Stars \citep{ORB6} are those of M dwarf systems within
25 pc.  About one third of these have periods longer than 100 yr, and
another third have periods 10--100 yr, with the remaining third
shorter than 10 yr.  Our SOAR program targeting orbits 0--6 yr is thus
well poised to make a significant contribution to this catalog.  All
but one of the orbits presented here have $P_\mathrm{orb}$ in that
range, demonstrating proof of concept for this plan.



Four of the five orbits presented here are new, while the fifth (LHS
501 AC) represents a substantial revision over the previously
published result \citep{Bar18}.  This set of orbit results is roughly
representative of the expected yield of our program: for most orbits,
we will combine existing data with our new data to produce fits for
systems that previously had no published orbits.  Several dozen more
orbits will be updates to systems that already had solutions
published; these will be a substantial fraction of the 73 targets in
our sample that already have orbits in the literature listed with $P_\mathrm{orb} \leq 30$ yr.

Overall, we expect to fit at least $\sim$50 orbits using the full
three years of observations planned for this program.  This estimate
is based on the number of systems already showing substantial motion
over the first 1.5 yr of observations and includes improvement of
published orbits as well as new orbits.  These will substantially
contribute to the 120-orbit goal to establish the M dwarf
$P_\mathrm{orb}$ vs.\ $e$ distribution, supplementing the planned
contributions from RECONS astrometry and the literature.

\subsection{Implications for the RECONS Astrometry Subset (0.9\,m PB)}

Of the 120 systems observed from the 0.9\,m PB list, 59 companions
were detected using SOAR.  Among these, 22 (37\%) had not been
resolved previously.  The lower yield of this subsample compared to
the other two is not too surprising because astrometry and speckle
interferometry are each somewhat sensitive to different types of
companions.  Speckle searches are most sensitive to equal-luminosity
components, but those systems exhibit no astrometric perturbation if both components
are equal luminosity and have the same mass.  In addition, many
of the astrometric companions are likely very low mass stars or brown
dwarfs that are beyond the magnitude difference limits of the speckle
observations.


When no companion is detected with speckle, the magnitude limits
reached at various separations constrain the nature of the astrometric
companion and its orbit.  Many cases in which these estimated mass
limits were notable are described in detail in the Appendix
($\S$\ref{sec:objectsofnote}).  For each system, we have used the
combined magnitude of the pair and the limiting magnitudes in the
speckle results to estimate the components' fluxes, which we then
combine with the size of the astrometric perturbation to estimate a
limit for each companion mass \citep[following][]{vdK67}.  Most of
these systems have been described in previous work in \textit{The
  Solar Neighborhood} series, often with plots showing their perturbed
astrometric residuals, hence our descriptions here can be considered
updates to those notes.

We used a similar procedure to estimate companion masses for the
0.9\,m PB systems that SOAR did resolve.  These masses (given in
$\S$\ref{sec:objectsofnote}) are only rough estimates determined from
the sizes of the photocentric displacements in the astrometric
pertubations, rather than the fully characterized photocentric orbits.
Future work will determine reliable photocentric orbits that can be
combined with these SOAR resolutions to yield dynamical masses for the
individual components.

\subsection{Implications for the Known Literature Multiples Subset}

Of the literature multiples, 140 out of 188 pairs observed were
resolved at SOAR.  These resolutions were the first ever for 34 of
these systems, while 106 had been previously resolved by others.  
Of the 48 unresolved systems, nearly all were initially identified as 
multiples through $JHK$ imaging or spectroscopic studies, hence their 
companions were likely too faint (e.g., brown dwarfs) or too closely 
bound (spectroscopic binaries) to resolve at SOAR in $I$ band. 
The new resolutions are systems 
that often have complementary (non-imaging) data in the literature,
and the previously-resolved systems have imaging that precedes our
SOAR results.  Both cases will assist in our orbit fitting goals, as
this extra information or lengthened time baseline both enable orbit
fits to be made earlier than with our SOAR data alone.  The five
orbits presented here demonstrate that concept.


Our selected 30-year orbital period limit is meant to capture as broad
a picture of M dwarf orbits as is feasible 
for a single observational program, 
in particular showcasing systems that fall 
between the widest binaries and those that are tightly bound because of tides. 
Inevitably, some of the
systems we have resolved will prove to have orbits longer than our
planned 30-year limit, as many of these are wider pairs initially
detected with less sensitive instruments.  In these cases, the data
from this campaign can be used to place constraints on those orbits,
e.g., choosing appropriate cadences on these slow-moving systems to
focus observations at future epochs when a companion moves quickly
through its periastron passage.  The ultimate contribution of
multi-decade orbits will thus come through observations collected over
multiple projects, with updated orbits more precisely determined than
currently possible.  For now, our SOAR observations of these
literature multiples provide a legacy dataset that will contribute to
future efforts long after our project is complete.

Long orbits can often be constrained by comparing their motions measured at two widely separated moments \citep[e.g.,][]{Bra19, Cur20, Bow21}. 
These recent efforts use the \textit{Hipparcos}-\textit{Gaia} Catalog of Accelerations \citep[HGCA;][]{Bra21}, which has presented recalibrated proper motions of systems measured by the \textit{Hipparcos} and \textit{Gaia} missions, $\sim$30 years apart. 
Another catalog based on the same principle has been compiled by \citet{Ker19}.
There are 60 systems on our program with an entry in the HCGA and many of these likely have accelerations evident in that catalog.
 By combining positional measurements with these proper motion changes, we could better constrain the orbits of these systems, especially those with very long periods. We will consider the use of this approach in our future work on orbits.

\subsection{Checking Criteria for Unresolved Multiples in Gaia DR2}
\label{sec:gaiaanalysis}


A total of 249 of the 252 systems were observed from the DR2 suspects
sample, selected at least partly based on their \textit{Gaia} DR2
astrometric fits (91 stars were included based {\it only} on those
fits).  This subset was created because during the SOAR sample
construction, the then-preliminary results of the \citet{Vri20}
analysis showed specific DR2 parameters to be reliable markers for
unresolved multiplicity.  The SOAR observations validate the defined
markers, with companions detected for 188 stars (76\% of that group).  Many of these
systems had no previously published resolved companions and are marked
with ** in column~2 of Table~\ref{tab:table_targets_results}.

The final analysis of \citet{Vri20} ultimately listed five criteria
that could be used to flag likely multiples in DR2 (given explicitly in $\S$\ref{sec:gai_sample}): missing parallax
or missing DR2 entry, and four threshold values of the DR2 astrometric
fit parameters.  That work involved constructing a sample of 542
RECONS parallax program targets that were cross-matched with
\textit{Gaia} DR2 results, and used those targets' multiplicity
information to identify the DR2 astrometric fit parameters that best
indicated the presence of unresolved companions.  For each of these four identified 
parameters, threshold values were then determined, above which three out of
four systems were unresolved multiples.  

Of the 252 systems in our
sample flagged in the preliminary stages of that DR2 analysis, 217 of
the stars observed fulfill two or more of the final \citet{Vri20}
criteria.  SOAR detected companions for 176 (81\%) of these 217
targets, confirming that the majority of poor fits in DR2 were due to
companions bright enough to detect with SOAR's HRCam+SAM.  As
\textit{Gaia}'s observing time baseline increases with future data
releases, these fit flags will reveal multiples with longer orbital
periods and fainter companions (smaller masses), as long as the
\textit{Gaia} data are fit with the single-star astrometric model.
Clearly, \textit{Gaia} data can be used to reveal many new potential
stellar multiples before the final release of its binary star
solutions.

For the 41 of 217 observed systems that fulfilled at least two of the
criteria from \citet{Vri20} but did not have a companion resolved, the
presence of a companion cannot be fully ruled out.  Indeed, roughly
half of this subset have had their companions already confirmed
through other means, such as spectroscopically or by showing unambiguous
orbital motion in RECONS astrometry.  Companions that are very faint
or orbiting close to their primary stars will not be detectable with
HRCam+SAM at SOAR; the largest magnitude difference observed here was
$\Delta I = 5.0$ and the smallest separation seen was 24 mas.  The DR2
suspects marked unresolved in Table~\ref{tab:table_targets_lookup}
must still be regarded as likely multiples, and future observations
are warranted to probe for very faint and very close-in companions.

To update the criteria for unresolved multiplicity of \textit{Gaia}
DR2 targets, we have added the new SOAR detections to the sample used
in \citet{Vri20}.  Although the sample used in that analysis was not
volume-complete beyond 13 pc, its proportion of multiples within any
distance matched the observed multiplicity found by more comprehensive
surveys \citep{Win19}.  To preserve that feature and avoid
overreporting multiples, we have updated the sample with these new
detections by only updating the multiplicity information for the
existing targets, without adding to that sample any new targets that
may have been observed here.  This sample multiplicity update does not
substantially change the \citet{Vri20} results.  The threshold values
of the four useful DR2 parameters may be lowered by $\sim$10\% to
select samples in which three out of four systems are unresolved
multiples.  The fifth criterion of missing DR2 entry or parallax
remains valid.  This consistency speaks to the robustness of the
overall results of \citet{Vri20}.


%
\section{Conclusions} 
\label{sec:conclusion}
%

In this work we have presented observations from the first 1.5 yr of
our planned 3-year speckle interferometry campaign at SOAR to observe
M dwarfs within 25 pc.  Key results to date include:

\begin{itemize}

\item speckle measurements of 333 M dwarfs in 320 systems; 211 (63\%)
  of these M dwarfs were resolved

\item four new orbits and one revised orbit with periods of 0.67--29
  yr for M dwarfs with masses of 0.15--0.30 M$_\odot$

\item measurements of resolved companions for 76\% 
  of candidate multiples from \textit{Gaia} DR2 identified
  by criteria for their astrometric fit parameters, as described in
  \cite{Vri20}
\end{itemize}

Each observation reported here of a stellar companion is a step toward
our goal of mapping the orbits of nearby M dwarf multiples.  Our
project specifically targets M dwarf systems with orbital periods of
0--30 yr and semimajor axes 0--6 AU and the five orbits presented here
span this full range, including some of the fastest-orbiting in our
sample and some with the richest sets of similar observations in the
literature.  Many systems had already been observed at SOAR prior to
this project and have measurements described in recent papers
\citep[e.g.,][]{Tok20,Tok21}.  HRCam+SAM at SOAR has had many
successful years observing stellar multiples \citep[10 yr as
  of][]{Tok18b}, and by focusing on the lowest-mass stars here we have
thoroughly demonstrated its capabilities regarding faint, red systems.

Since the preparation of this paper began, with each observing run we
have noted more systems that have enough data for orbit fits.  This
speckle program is thus well on its way to forming a significant
contribution to the overall project of mapping M dwarf orbits, and we
anticipate continued success in the remaining 1.5 yr of this program.
A future publication at the conclusion of this campaign will include
several times the number of orbits presented here.

This project is an effort bringing together several observing methods,
and as such demonstrates the power of these methods to complement and
inform each other.  Long-term ground-based astrometry from RECONS
provides many full orbits and highlights systems with anomalous motion
(but not necessarily distinguishable orbits) for speckle follow-up.
The speckle interferometry from SOAR confirms or constrains those
systems, and also efficiently captures the equal-mass systems that are
not easily detectable via unresolved astrometry.  Speckle observations
may be combined with other resolutions in the literature, e.g., from
adaptive optics, allowing orbits to be observed and characterized over
long time baselines.

A multi-method approach is essential to this project, as the spatial
scales involved in binary star formation and dynamical evolution span
orders of magnitude in AU.  The complex mix of physics may depend on
several fundamental properties such as mass, system mass ratio, and
age, making it imperative that a wide range of orbits be considered to make 
meaningful comparison between models and observations.
Ultimately, the multiples reported here have far-reaching potential
consequences for M dwarf multiplicity, star formation, and local
Galaxy mass distribution.  This is because M dwarfs dominate the
Galactic population, accounting for three out of every four stars
\citep{Hen06,Hen18}.  It is therefore essential to use all of the
observing techniques at our disposal to determine not only which
systems have companions, but to measure accurate sizes and shapes for
their orbits, as those clues will reveal how the systems formed.


\acknowledgments

{Colleagues at the Southern Astrophysical Research (SOAR) Telescope,
  the Cerro Tololo Inter-American Observatory (CTIO), and the SMARTS
  Consortium and have made this work possible.

This work has made use of data from the European Space Agency (ESA)
mission {\it Gaia} (\url{https://www.cosmos.esa.int/gaia}), processed
by the {\it Gaia} Data Processing and Analysis Consortium (DPAC,
\url{https://www.cosmos.esa.int/web/gaia/dpac/consortium}). Funding
for the DPAC has been provided by national institutions, in particular
the institutions participating in the {\it Gaia} Multilateral
Agreement.

This research has made use of the Washington Double Star Catalog
maintained at the U.S. Naval Observatory.

The National Science Foundation has been consistently supportive of
this effort under grants AST-0507711, AST-0908402, AST-1109445,
AST-141206, AST-1715551, and AST-2108373.



}

\facilities{CTIO:0.9\,m, CTIO:SOAR}

\appendix
%
\section{Systems worthy of note}
\label{sec:objectsofnote}
%

Here we describe several systems for which these SOAR results add
significant new information or shed light on unusual observational
histories.  They are listed in order of ascending R.A., with WDS
coordinate designations given in parentheses.  The RECONS astrometry
program mentioned for many systems refers to the long-term effort at
the SMARTS 0.9\,m.






%
%
%
%
%
%
%




\begin{itemize}

\item SCR 0128-1458 AB (01287$-$1458):  

Through four resolutions at SOAR, we have confirmed the presence of
this companion first noted tentatively in RECONS astrometry residuals
by \citet{Rie18}.  The $\Delta I$ of 2.6--2.7 mag indicates the
companion has mass $\lesssim$0.2 M$_\odot$.  Continuing observations
will provide valuable future constraints for the photocentric orbit in
the RECONS astrometry, which is still incomplete after 10 yr of data.

\item LEHPM 1-1882 AB (01477$-$4836):  

\citet{Win17} revealed this binary via RECONS astrometry residuals.
Its period is long, with the orbit not yet complete in what is now 15
yr of data.  Although \citet{Win17} suggested the secondary companion
contributes little light in $R$ band, our three SOAR resolutions at
$I$ indicate a stellar companion with luminosity similar to the
primary.


\item LHS 1561 AB (03347$-$0451):  

Seven SOAR observations over 2018.8--2020.9 have resolved this
system's secondary to have moved 20$^\circ$ through its orbit.
\citet{Jef18} reported this system to be a spectroscopic triple; the
tertiary is presumably less luminous and/or more closely bound to the
primary, as our observed component's motion and $\Delta I$ indicate
that we are consistently resolving the same (secondary) companion.


\item LHS 1582 AB (03434$-$0934): 

This system's 5 yr photocentric orbit was fully characterized in
\citet{Vri20}, but the companion was not detected in our two SOAR
observations.  Comparison of the photometric (13 pc) and trigonometric
(20 pc) distances by \citet{Rie10} and \citet{Lur14} indicated that
the companion contributes noticeable light to the system.  The
limiting $\Delta I$ values of 1.4 mag at $0\farcs15$ and 4.3 mag at
$1\farcs0$ from SOAR suggest it has mass $\lesssim \!0.15$
M$_\odot$.

\item GJ 1068 (04105$-$5336): 

Two observations of this target revealed a relatively closely
separated background star; at 2019.6136 its separation and position
angle were $3\farcs7177$ and 38.6$^\circ$, and at 2020.1111 they were
$5\farcs1628$ and 35.5$^\circ$.  Comparison with archival images from
the CTIO/SMARTS 0.9\,m confirm that this background star is not bound
to GJ 1068.  This target's results are thus not included in
Table~\ref{tab:table_targets_results}.

\item SCR 0702-6102 AB (07028$-$6103):  

We identified this system's companion early in the SOAR program as a
fast mover, and have resolved it seven times from 2019.86--2020.99.
 The companion creates a low-amplitude
perturbation in the RECONS astrometry residuals \citep[as noted
  in][]{Win17} with a period of $\sim$2.5 yr.  That motion is
consistent with what we have observed in the SOAR data.

\item SCR 0723-8015 AB (07240$-$8015):

This system's color and absolute magnitude are consistent with a
$\sim$0.1 M$_\odot$ star, and the clear perturbation indicates an
orbital period that has not yet wrapped in 17 years of RECONS
astrometry data.  The companion has not been detected in three
observations in $I$ at SOAR to limits of $\Delta I$ = 1.6 and 3.0 at
$0\farcs15$ and $1\farcs$, respectively, indicating that it is of
very low mass.  This implies that the companion is a very low
luminosity red or white dwarf, or a brown dwarf.

\item SCR 0838-5855 AB (08380$-$5856):  

The RECONS astrometry indicates a large perturbation first shown in
\citet{Win17} that now exceeds 50 mas in both RA and Decl.~directions,
but has not wrapped in 14 years of coverage.  The two new SOAR
resolutions are the first ever for this system and indicate the
companion has $M_I$ = 14.6, placing it very near end of the main
sequence with a mass $\lesssim$0.1 M$_\odot$.

\item LHS 2071 AB (08553$-$2352):  

This system was first noted as binary by \citet{Rie10}, who presented
a preliminary fit to the partially observed orbit in RECONS astrometry
data.  Ten additional years of data have revealed the orbital period
to be greater than the 21 years of coverage to date.  The four SOAR
observations show clear orbital motion from 2018.2--2020.8; these will
allow us to constrain the incomplete photocentric orbit in future
work.  The consistent $\Delta I$ of 2.4--2.6 mag indicate the
companion has mass $\lesssim$0.2 M$_\odot$, but it is not substellar.

\item LP 788-001 AB (09314$-$1718):   

\citet{Win17} showed clear orbital motion for this system in the
RECONS astrometry residuals, and noted that the companion must
contribute little flux in $I$ band.  The orbit has not wrapped after 8
years of coverage, and our SOAR observation in $I$ did not reveal the
companion.  Because the absolute magnitude of this system sets the
primary mass at $\sim$0.1 M$_\odot$, the detection limits suggest that
the companion is substellar.

\item LP 848-050 AB (10427$-$2416): 

This system exhibits an $\sim$8 yr orbit in the RECONS astrometry with
a large-amplitude photocentric perturbation \citep[see Figure 8
  of][]{Win17}.  Because the color and absolute magnitude of the
system are consistent with a $\sim$0.1 M$_\odot$ star, the two
non-resolutions at SOAR suggest that the companion is either a very
low luminosity red or white dwarf, or a brown dwarf.

\item L 327-121 AB (12336$-$4826):  

The RECONS astrometric perturbation for this system shown in
\citet{Win17} has continued in recent data now spanning 10 yr.  This
is likely the reason this system has a poor fit in \textit{Gaia} DR2
and no parallax given in EDR3.  The orbital period is $\sim$9 yr, and
a robust fit of this orbit will be possible in future work, enabling
dynamical masses to be determined by combining that fit with these
three new SOAR resolutions.  \citet{Win17} noted an excessive mismatch
between photometric and trigonometric distance, suggesting either that
the system is young or includes a third luminous component.  The SOAR
data indicate $M_I$ values of 8.43 and 8.83 for the two components,
implying masses of 0.4--0.5 M$_\odot$ and 0.3--0.4 M$_\odot$.  The mass sum is consistent with the
orbital information available, indicating that a third luminous
component is unlikely.



\item LTT 6288 (15457$-$4330):  

This system's photocentric orbit was first described in \citet{Win17}
and later updated in \citet{Vri20}.  The orbital period is 9.9 yr.
The two resolutions at SOAR indicate the companion's luminosity is
consistent with mass $\lesssim$0.2 M$_\odot$, with the primary roughly
twice as massive.  The reliable RECONS astrometric orbit and
continuing SOAR observations will enable a precise dynamical mass
determination for both components in future work.

\item SCR 1546-5534 AB (15467$-$5535):   

The orbit shown in \citet{Hen18} has continued in now 9 years of
RECONS astrometry, with preliminary fits suggesting an orbital period
of $\sim$7 yr.  The two SOAR resolutions reveal the companion to be
somewhat less massive than the primary star, with the secondary's
absolute $I$ magnitude consistent with $\lesssim$0.1 M$_\odot$ and the
primary's consistent with roughly twice that mass.  The secondary is
more likely stellar than substellar, however, as \citet{Hen18} pointed
out overluminosity evident in the $\sim$30\% difference between
photometric (7 pc) and trigonometric (10 pc) distances for this
system.

\item LHS 3117 AB (15474$-$1054):  

\citet{Zec09} noted a radial velocity trend in VLT+UVES (Ultraviolet
and Visible Spectrometer) data over $\sim$500 d starting in 2004, and
noted this system as SB1.  This signal was confirmed by the
re-analysis of the same data by \citet{Tuo14}.  Our three new
observations at SOAR over 2019.5--2020.2 reveal the companion, and the
$\Delta I$ of 0.8--1.0 mag indicates it is likely a low-mass star
rather than a brown dwarf.

\item GJ 1212 AB (17137$-$0825): 

This system has been noted as a spectroscopic binary by \citet{Rei12},
\citet{Hou15}, and \citet{Jef18}.  
No relative positions have been published before our SOAR observations.  
These three resolutions show component
B moving quickly around A from 2019.5--2020.2, sweeping through
191$^\circ$ in position angle.  Estimating the orbital semimajor axis
to be 1--3 times the maximum displacement seen so far and assuming
mass sums of 0.5--0.7 M$_\odot$ yields orbital periods of 0.97--5.9
yr.  This target is thus high priority for continued observations and
orbit characterization on our SOAR program.

\item G 154-043 AB (18036$-$1859):  

Revealed as binary via the astrometric perturbation shown in
\citep{Win17}, 10 years of RECONS data now show this system to have an
orbital period of 8--12 yr.  The two observations at SOAR indicate
that this binary has components with $M_I$ = 10.57 and 11.92, implying
masses of 0.15 M$_\odot$ and 0.12 M$_\odot$.  The SOAR data also show significant motion
through 27$^\circ$, so future work should allow for a refined orbit
and reliable masses.

\item LTT 7434 AB (18460$-$2856):   

As highlighted in \citet{Win17}, this system has historically been
challenging to interpret.  The trigonometric distance is 1.4 times the
photometric distance, implying two equal-mass components, yet the
strong astrometric perturbation is only possible with unequal-mass
components.  Additional RECONS astrometry data acquired since
\citet{Win17} continues the perturbation shown there, with the orbital
period now estimated to be more than 20 yr.  At SOAR we have twice
resolved a companion at $0\farcs35$--$0\farcs39$ (2019.61--2020.77) 
that is 1.4 mag fainter than
the primary in $I$ band; these are the first resolutions of this
system.  \citet{Bon13} noted that this system is an SB2 with variable
line width, suggesting the possibility of a close third component that could explain the
excess flux.  We will continue monitoring the long-term astrometry to
complete the orbit and to look for any additional perturbations from a
potential third companion.

\item GJ 829 AB (21296$+$1739): 

\citet{Del99} first reported this system to be binary and
characterized its spectroscopic orbit.  It was reported as visually
resolved by \citet{Opp01} at Palomar and by \citet{Die12} with
HST/NICMOS, but in both cases no details of the resolutions are given.
Our SOAR observations of the companion at 25.0--36.7 mas separations
are the most detailed to date.  The close separation of this system
presents a challenge for HRCam+SAM to resolve consistently, but its
53-day orbital period \citep{Del99} give us ample future opportunities
to attempt observations.  When we have observed the entire orbit
visually, fitting that data will yield the orbital inclination, which
we will combine with the \citet{Del99} spectroscopic fit to obtain the
individual component masses.

\item LHS 3739 BC (21588$-$3226, a.k.a.\ LHS 3738 AB): 

The A-BC separation is $113^{\prime\prime}$, forming a hierarchical
triple.  \citet{Rie10} first announced the companion to B based on
RECONS astrometry and noted no significant overluminosity, indicating
a much lower-mass companion.  \citet{Lur14} presented an updated
photocentric orbit; the six additional years of RECONS astrometry
since then are consistent with that result.  The BC pair has not been
resolved at SOAR in two attempts, with limits of $\Delta I$ = 2.3 at
$0\farcs15$ and 3.4 at $1\farcs0$, implying a companion with mass
lower than $\sim$0.1 M$_\odot$.

\item LEHPM 1-4771 (22302$-$5345):  

Although this binary's orbital motion was shown in \citet{Win17} and
its orbit fit updated in \citet{Vri20}, the five SOAR observations
reported here represent the first resolutions of the pair.  The
magnitude difference of $\Delta I$ = 0.9--1.2 mag indicates a
secondary somewhat less massive than the primary, consistent with the
assertion in \citet{Win17} that the secondary must contribute little
flux in the $R$ band.  Once more of the $\sim$6 yr orbit is covered
with SOAR observations, we will combine the photocentric fit with SOAR
resolutions to determine dynamical masses for the components.

\item LTT 9084 AB (22351$-$4218):  

This system was first resolved by \citet{Kar20} in July 2013, who
found the binary to be separated by 398--405 mas with position angle
333$^\circ$--334$^\circ$, and brightness differences of $<$0.2 mag in
each of $JHK$ bands.  Our SOAR observations yield $\Delta I$ = 0.0,
consistent with the near-infrared values.  Thus, the components are
likely of similar mass.  
Our observations spanning 2019.5--2020.8 show 
the secondary moving from 428 mas to 401 mas, 
to nearly the same separation as 
observed in 2013 by \citet{Kar20}.  
The position angles we observed,
however, were 17$^\circ$--21$^\circ$ greater than the 2013
observations, increasing through 2019.5--2020.8.  This displacement
suggests the companion passed through due north in the 6 yr between
2013 and 2019.  Together, the available data suggest the orbit is
either highly inclined or highly eccentric.  Although the orbit is
likely several decades in duration, continued observations over the
next two years could rule out one of the above scenarios through any
variations in the secondary's speed.

\end{itemize}


\end{document}